\documentclass[11pt,oneside]{article}

\usepackage{amsthm}
\usepackage{amsmath}
\usepackage{amssymb}
\usepackage{graphicx}
\usepackage{color}
\usepackage{dsfont}
\usepackage[normalem]{ulem}

\newtheorem{theorem}{Theorem}[section]

\newtheorem{proposition}[theorem]{Proposition}

\theoremstyle{definition}
\newtheorem{definition}[theorem]{Definition}
\newtheorem{example}[theorem]{Example}
\theoremstyle{remark}
\newtheorem{remark}[theorem]{Remark}

\numberwithin{equation}{section}

\def\cD{\mathcal{D}}
\def\cF{\mathcal{F}}
\def\cT{\mathcal T}
\def\cQ{\mathcal{Q}}
\def\cU{\mathcal{U}}
\def\cP{\mathcal{P}}
\def\bP{\mathbb{P}}

\usepackage{url}
\makeatletter\def\url@leostyle{%
  \@ifundefined{selectfont}{\def\UrlFont{\sf}}{\def\UrlFont{\scriptsize\ttfamily}}} \makeatother

\title{Dynamic Coherent Acceptability Indices and their \\ Applications to Finance
\thanks{TRB and IC acknowledge support from the NSF grant DMS-0908099. The authors would like to thank the anonymous referees and the editors for their helpful comments and suggestions which improved greatly the final manuscript.}}

\author{
Tomasz R. Bielecki \\
\url{bielecki@iit.edu}
\and
Igor Cialenco \\
\url{igor@math.iit.edu}
\and
Zhao Zhang \\
\url{zzhang35@iit.edu}
}

\date{
{\small  Department of Applied Mathematics \\[-0.3ex]
         Illinois Institute of Technology \\[-0.3ex]
         10 West 32nd Str, Bld E1, Room 208 \\[-0.3ex]
         Chicago, IL 60616-3793}  \\[0.7ex]
 First circulated: October 21, 2010\\
 Current version: \today \\[0.7ex]
}

\begin{document}

\maketitle

\begin{abstract}
In this paper we present a theoretical framework for studying coherent acceptability
indices in a  dynamic setup. We study  dynamic coherent acceptability indices and dynamic coherent risk measures, and we establish a duality between them. We derive a representation theorem for dynamic coherent risk measures in terms of so called dynamically consistent sequence of sets of probability measures. Based on these results, we give a specific    construction of dynamic coherent acceptability indices. We also provide examples of dynamic coherent acceptability indices, both abstract and also some that generalize selected classical financial measures of portfolio performance.
\vskip 0.3 true cm
\end{abstract}

{\noindent \small
}



\section{Introduction}
\noindent
Individual and institutional investors are typically concerned with finding satisfactory balance between \textit{reward and risk} associated with an investment process. Various measures have been developed to quantify this balance. Such measures are typically referred to as \textit{performance measures} or \textit{measures of performance} (MOP). Recently, Cherny and Madan \cite{ChernyMadan2009} originated an effort to provide a mathematical framework to study these measures in a unified way. The present paper contributes to this effort.

One of the most popular MOPs is the Sharpe Ratio (SR) introduced in \cite{Sharpe1966}. SR is expressed as a \textit{ratio} of expected excess return to standard deviation, and thus in financial applications it measures expected excess return of a portfolio in units of portfolio's standard deviation. SR has been used as a classical tool to rank portfolios according to their ``reward-to-risk" characteristics.

Using standard deviation to quantify risk is considered to be the major drawback of Sharpe Ratio. The reason of course is that positive returns also contribute to this measure of risk. To eliminate this unwanted feature other ratio-types MOPs were proposed, such as Sortino Ratio (SOR) \cite{SortinoPrice1994} and Gain Loss Ratio (GLR) \cite{GLR2000}. These MOPs focus on downside risk only.
A popular generalization of SR is provided by the Risk Adjusted Return on Capital (RAROC), which is constructed as a  ratio of mean excess return to some selected measure of risk.

All the MOPs mentioned above share some common desirable features: they are unit-less, they are increasing functions of reward and decreasing functions of risk; moreover, according to these MOPs diversification of a portfolio improves its performance.
This observation prompts a natural desire to study MOPs in a unified mathematical framework.\footnote{There exists a vast literature that studies measures of risk
in a general mathematical framework.} As already mentioned, such a study was recently originated in \cite{ChernyMadan2009}.
We shall recall the main results of that paper in Section \ref{sec:SAI} below. The study of \cite{ChernyMadan2009} was done in static, one-time period setup. Cherny and Madan coined the term \textbf{Acceptability Index} (AI) as a mathematical terminology for MOPs. Our goal is to elevate the mathematical framework for studying AIs to dynamical, multi-period setup, where cash flows are considered as random processes, and one needs to assess their acceptability consistently in time. In particular, we are concerned not just with the total cumulative terminal value of the cashflow as seen from the initial time of the investment process, but also with all remaining cumulative cashflows between each intermediate time and the terminal time of the investment process.

Thus, in a sense, our program is analogous to the one of those researchers
(cf. \cite{ArtzerDelbaenEberHeathKu2007, BionNadal2009, CheriditoDelbaenKupper2004, CheriditoDelbaenKupper2005, CheriditoDelbaenKupper2006,
CvitanicKaratzas1999, Detlefsen2005, JobertRogers2006, Riedel2004,
FrittelliScandolo2006, FrittelliGianin2004, Roorda2005, Weber2006, DelbaenPengGianin2010, Tutsch2008}) who are studying dynamic risk measures.
Moreover, as it will be seen later in the paper, there is a duality relationship between dynamic (coherent) acceptability indices and dynamic (coherent) risk measures in the sense of Section \ref{sec:DCRM}.

\medskip

\noindent
The paper is organized as follows:

In Section \ref{sec:SAI} we summarize the main results of \cite{ChernyMadan2009}. This is done for the convenience of the reader, but also in order to give the flavor of the duality between acceptability indices and risk measures, that will be generalized to the dynamic framework in the subsequent sections. In Section \ref{sec:DAI} we present the  definition of a  dynamic coherent  acceptability index (DCAI). We devote some time to discussion of the properties of DCAI from the definition, putting special emphasis on discussion of various forms of the dynamic consistency property.

Section \ref{sec:DCRM} first introduces the concept of the  dynamic coherent risk measure (DCRM), specific for our needs, and then proceeds to study the duality between families of such measures and DCAI. In the process, we discuss the dynamic consistency property of a DCRM, and we relate our findings to the results from existing literature. 

In Section \ref{sec:SCCDAI} we provide characterization of a DCRM in terms of so called dynamically consistent sequence of sets of probability measures, thereby providing an additional perspective at DCAIs.

Section \ref{sec:EX} is dedicated to discussion of some abstract examples of dynamic MOPs, as well as some specific examples of dynamic MOPs
derived form the classical ones, such as GLR and RAROC. In particular, we show that dynamic version of GLR is a DCAI, whereas the dynamic version of RAROC is not.

\section{Static Acceptability Indices}\label{sec:SAI}

In this section, we will briefly review the theory of static acceptability indices developed   in \cite{ChernyMadan2009}.

Let $(\Omega, \mathcal{F}, \mathbb{P})$ be a probability space and denote by $L^\infty(\Omega, \mathcal{F}, \mathbb{P})$ the space of all
bounded random variables on $(\Omega, \mathcal{F}, \mathbb{P})$. The random variable $X\in L^\infty$ can be regarded as discounted {\it terminal} cash flow of a zero-cost self-financed portfolio. By definition,
an acceptability index is a map $\alpha : L^\infty\to [0,+\infty]$.
The   value $\alpha(X)$  should be understood as the degree of acceptability of a cash flow
 $X$; in a sense, it  represents a measure of efficiency of the cash flow. A larger index indicates
better performance, with $\alpha(X)=+\infty$ for $X$ being an `arbitrage opportunity'; in particular, if the cash flow is strictly positive, then $\alpha(X)=+\infty.$

Acceptability index as such is a too broad concept, and it may not fulfill certain practically desirable properties. That is why, Cherny and Madan \cite{ChernyMadan2009} focused their attention on a more specific concept of the coherent acceptability index.

\begin{definition}
An acceptability index is called {\it coherent} if the following properties are satisfied:
\begin{enumerate}
\item[(S1)] {\it Monotonicity.} If $X\leq Y$, then $\alpha(X) \leq \alpha(Y)$;

\item[(S2)] {\it Scale invariance.}
For every $X\in L^\infty$ and $\lambda > 0, \ \alpha(\lambda X) = \alpha(X)$;

\item[(S3)] {\it Quasi-concavity.} If $\alpha(X) \geq x, \ \alpha(Y)\geq x$, then
$ \alpha(\lambda X + (1-\lambda)Y) \geq x$ for all $\lambda\in[0,1]$;

\item[(S4)] {\it Fatou property.}
If $|X_n|\leq 1, \ \alpha(X_n)\geq x$ for all $n\geq 1$,  and $X_n\to X$, as $n\to\infty$, in probability, then $\alpha(X)\geq x$.
\end{enumerate}
\end{definition}

The above properties have natural financial interpretation. For example, (S1) states that if $Y$ dominates $X$ -- $\mathbb{P}$ almost surely,\footnote{In the present paper we shall make a standing assumption that $\Omega$ is finite and that $\mathbb{P}$ is strictly positive. Thus, our statements regarding relations between random variables will hold point-wise. In particular, $Y\geq X$ will mean that  $Y$ dominates $X$ for every $\omega \in \Omega.$} then $Y$ is acceptable at least at the same level as $X$ is; (S2) implies that cash flows with the same direction of trade have the same level of acceptance. Quasi-concavity (S3) implies that a diversified portfolio performs at higher level than its components; to see this, it is enough to take $x=\min\{\alpha(X),\alpha(Y)\}$. Fatou property (S4) is a technical continuity property, which is
used for constructing the duality between coherent acceptability indices and coherent risk measures.

It can be shown that Sharpe Ratio, defined as $\mathrm{SR}(X):= \frac{\mathbf{E}(X)-r}{{\rm STD }(X)}$, where ${\rm STD}(X)$ is the standard deviation of $X$ and $r$ is the (constant) interest rate,  does not satisfy the monotonicity property (S1), and hence it is not a coherent acceptability index. The
Gain Loss Ratio, given by $\mathrm{GLR}(X):=\mathbb{E}(X)/\mathbb{E}(\max\{-X,0\})$ if
${X}>0$, and zero otherwise, is a coherent acceptability index. Other measures of performance such as
RAROC, AIT, AIW, AIMIN, AIMAX, AIMINMAX, AIMAXMIN etc, have been also studied in \cite{ChernyMadan2009}.
Moreover, in \cite{ChernyMadan2009} the authors proved the following representation theorem:

\begin{theorem} \label{th-AI-static-Chernyi}
A map $\alpha: L^\infty\to [0,+\infty]$, unbounded from above, is a coherent acceptability index
if and only if there exists a family $(\mathcal{D}_x)_{x\in[0,+\infty]}$ of sets of
 probability measures,
such that $\mathcal{D}_x\subset\mathcal{D}_y$ for $x\leq y,$ and $\alpha$ admits the following representation
\begin{equation} \label{eq:repStaticCAI}
\alpha(X) = \sup \left\{  x\in[0,+\infty) \ : \ \inf\limits_{Q\in\mathcal{D}_x} \mathbb{E}_Q[X] \geq 0 \right\} \ ,
\end{equation}
where $\inf\emptyset = \infty$ and $\sup\emptyset=0$.
\end{theorem}
Thus, any Coherent Acceptability Index (CAI) can be characterized by an increasing family of sets of probability measures.
This family of probability measures can be seen as generalized scenarios as described in \cite{ADEH1999},
or set of supporting kernels as discussed in \cite{ChernyMadan2009}.
Moreover, there is a strong relationship between CAI and
Coherent Risk Measures (CRM), a concept introduced by Artzner, Delbaen, Eber, Heath \cite{ADEH1997, ADEH1999}.

\begin{definition}\label{d23}
A function $\rho:L^\infty\to\mathbb{R}$ is called {\it coherent risk measure} if the following properties are satisfied:
\begin{enumerate}
\item[(R1)] {\it Monotonicity.} If $X\leq Y$, then $\rho(X) \geq \rho(Y)$;

\item[(R2)] {\it Positive homogeneity.}
$\rho(\lambda X) = \lambda\rho(X)$, for every $X\in L^\infty$ and $\lambda \geq 0$;

\item[(R3)] {\it Translation property.} $\rho(X+k) = \rho(X) - k$, for every $X\in L^\infty$ and $k\in\mathbb{R}$;
\item[(R4)] {\it Subadditivity.} $\rho (X+Y) \leq \rho(X) + \rho(Y)$, for every $X,Y\in L^\infty$.
\end{enumerate}
\end{definition}

Traditional Value at Risk
$\mathrm{VaR}_\alpha(X):=\inf\{ m\in\mathbb{R}  \  | \ \mathbb{P}[X+m < 0]\leq \alpha\}$,
while very popular, it is not a coherent risk measure since it lacks the
subadditivity property (R4), which corresponds to the diversification property in finance. In contrast, the Tail Value at Risk (also called Tail Conditional Expectation), defined as
$\mathrm{TVaR}_\alpha(X) : = -\inf_{\mathbb{Q}\in{\mathcal Q}_\alpha}\mathbb{E}_{\mathbb{Q}}[X ]$, where $\alpha\in(0,1]$ and ${\mathcal Q}_\alpha$ is the set of probability measures absolutely continuous with respect to $\mathbb{P}$ such that $d\mathbb{Q}/d\mathbb{P}\leq \alpha^{-1}$, is a CRM. So is the
Weighted Value at Risk, $WVaR_{\mu}(X):=\int_{(0,1]}TVaR_\alpha(X)\mu(d\alpha)$, where $\mu$ is a probability measure on $(0,1]$. The following representation
theorem is established in \cite{ADEH1999} for finite $\Omega$, and generalized  to a general probability space in \cite{Delbaen2002, CarGemanMadan2001}:

\begin{theorem}\label{th:ReprThStaticCRM}
A function $\rho:L^\infty\to\mathbb{R}$ is a coherent risk measure if and only if
\begin{equation}\label{eq:repTh-staticRM}
\rho(X) = \sup \{\, \mathbb{E}_\mathbb{\mathbb{Q}}[-X] \ : \  \mathbb{Q}\in\mathcal{P} \} \,
\end{equation}
for a certain set $\mathcal{D}$ of probability measures absolutely continuous with respect to $\mathbb{P}$.
\end{theorem}
Note that by \eqref{eq:repStaticCAI} and the above theorem, every CAI can be characterized in terms of an increasing
family of coherent risk measures.

The theory of static risk measures has been explored and extended
by many researchers; to mention just a few of them:
F\"ollmer and Schied \cite{FollmerSchied2002a, FollmerSchied2002b} and Frittelli and Rosazza Gianin \cite{FrittelliGianin2002}  generalized the concept of
coherent risk measures to convex and monetary risk measures; Cheridito and Li \cite{CheriditoLi2009} studied generalized measures on Orlicz Hearts,
law-invariant risk measures have been investigated by Kusuoka \cite{Kusuoka2001} and Frittelli and Rosazza Gianin \cite{FrittelliGianin2005};
for a systematic discussion on static risk measures we refer reader to the monographs by Delbaen \cite{Delbaen2000} and
F\"ollmer and Schied \cite[Chapter 4]{FollmerSchiedBook2004}.

\section{Dynamic Coherent Acceptability Indices}\label{sec:DAI}

As it has been already stated, the dynamic acceptability indices are meant to assess performance of a cash-flow accounting for newly acquired information when time progresses. Of course, one may attempt to use for this purpose a sequence of static (one-period) acceptability indices. However, by doing this one may end up with a sequence of measurements that are not consistent in time, in the sense to be explained below (cf. property D7). The motivation for developing a theory of DCAIs, as presented in this paper, was to create a systematic mathematical framework to provide performance measurements consistently in time.

Let $(\Omega, \mathcal{F}, \mathbb{P})$ be a finite underlying probability space, and let $\mathcal{T}=\{0,1,2,\ldots,T\}$ be a finite set of time instants.
We assume that $\mathbb{P}$ is of full support. We endow the underlying probability space with a filtration $\mathbb{F} = \{\mathcal{F}_t\}_{t=0}^T$. For each $t\in \cT$ and $\mathcal{F}_t\in\mathbb{F}$, there exists a partition of $\Omega$, say $\{P^t_1,P^t_2,\dots,P^t_{n_t}\}$, that generates $\mathcal{F}_t$.

A cash flow, also called dividend process, denoted as $D=\{D_t(\omega)\}_{t=0}^T$, is any real valued random process adapted to the filtration $\mathbb{F}$. We denote by $\mathcal{D}$ the set of all cash flows. In addition, we denote by $\mathcal{P}$ the set of all probability measures that are absolutely continuous  with respect to $\mathbb{P}$, and by $\mathcal{P}^e$ the set of all probability measures equivalent to $\mathbb{P}$. Also,  $c$ will denote a generic constant, and $m$ will denote a generic  random variable. Finally, a standing (financial type) assumption, which we make without loss of generality, is that the interest rates are zero.

\begin{definition}
A dynamic coherent acceptability index is a function
$\alpha: \cT\times\mathcal{D}\times\Omega\to [0,+\infty]$ that satisfies the following properties:
\begin{enumerate}
\item[\bf (D1)] {\bf Adaptiveness.}
For any $t\in\cT$ and $D\in\mathcal{D}$, $\alpha_t(D)$ is $\mathcal{F}_t$-measurable;
\item[\bf (D2)] {\bf Independence of the past.}
For any $t\in\cT$ and $D, D'\in\mathcal{D}$, if there exists $A\in\mathcal{F}_t$ such that $1_A D_s=1_A D'_s$ for
all $s\geq t$, then $1_A\alpha_t(D)=1_A\alpha_t(D')$;
\item[\bf (D3)] {\bf Monotonicity.}
For any $t\in\cT$ and $D, D'\in\mathcal{D}$, if $D_s(\omega)\geq D'_s(\omega)$ for all $s\geq t$ and $\omega\in\Omega$, then $\alpha_t(D,\omega)\geq \alpha_t(D',\omega)$ for all $\omega\in\Omega$;
\item[\bf (D4)] {\bf Scale invariance.}
$\alpha_t(\lambda D,\omega) = \alpha_t(D, \omega)$ for
all  $\lambda >0, \ D\in\mathcal{D}, \ t\in\cT,$ and $\omega\in\Omega$;
\item[\bf (D5)] {\bf Quasi-concavity.}
 If $\alpha_t(D, \omega) \geq x$ and $\alpha_t(D^\prime, \omega)\geq x$ for some $t\in\cT$, $\omega\in\Omega$,
 $D,D'\in\mathcal{D}$, and $x\in (0,+\infty]$, then
 $\alpha_t(\lambda D + (1-\lambda)D^\prime,\omega) \geq x$ for all $\lambda\in[0,1]$;
\item[\bf (D6)] {\bf Translation invariance.}
    $\alpha_t(D+m1_{\{t\}},\omega)=\alpha_t(D+m1_{\{s\}},\omega)$ for every
    $t\in\cT$, $D\in\mathcal{D}$, $\omega\in\Omega$, $s\geq t$ and every $\mathcal{F}_t$-measurable random variable $m$;
\item[\bf (D7)] {\bf Dynamic consistency.} For any $t\in[0,\ldots, T-1]$ and $D,D'\in\mathcal{D}$, if $D_t(\omega) \geq 0 \geq D'_t(\omega)$ for all $\omega\in\Omega$, and there exists a non-negative $\mathcal{F}_t$-measurable random variable $m$ such that $\alpha_{t+1}(D,\omega)\geq m(\omega)\geq \alpha_{t+1}(D',\omega)$ for all $\omega\in\Omega$, then $\alpha_t(D,\omega)\geq m(\omega)\geq \alpha_t(D',\omega)$ for all $\omega\in\Omega$.
\end{enumerate}
\end{definition}

\noindent Property (D1) is a natural property in a dynamic setup and it assumes that a DCAI is adapted to the same information flow $\{\mathcal{F}_t\}_{t\geq 0}$ as is any cash flow $D\in\mathcal{D}$. \smallskip

\noindent
 (D2) postulates that in the dynamic context the current measurement of performance of a cash flow $D$ only accounts for future payoffs. To decide, at any given point of time, whether one should hold on to a position generating the cash flow $D$, one may want to compare the measurement of the performance of the future payoffs (provided by DCAI at this  point of time) to already known past payoffs.
\smallskip

\noindent
Properties (D3)-(D5) are naturally inherited from the static case (cf. Section \ref{sec:SAI}).
\smallskip

\noindent
Translation invariance (D6) implies that  if a known dividend $m$ is added to $D$ at time $t$ (today), or at any future time $s\geq t$, then all such adjusted cashflows are accepted today at the same level.

\smallskip

\noindent
Dynamic consistency (D7) is the property in  the dynamic setup which relates the values of the index between two consecutive days in a consistent manner.
It can be interpreted from financial point of view as follows:
if a portfolio has a nonnegative cashflow today, then we accept this portfolio today at least at the same level as we would accept it tomorrow;
similarly, if the today's cashflow is nonpositive the acceptance level today can not be larger than the level of acceptance tomorrow.

For technical reasons, which will become clear later, we assume that
for every DCAI $\alpha$,  and for every $t\in\cT$ and $\omega\in\Omega$, there exist two portfolios $D,D'\in\mathcal{D}$
such that $\alpha_t(D,\omega)=+\infty$ and $\alpha_t(D',\omega)=0$. We shall say that DAI $\alpha$ is {\it normalized}.

Note that normalization will exclude the degenerate examples of acceptability indices such as a constant index over all states, times, and portfolios.
Moreover, one can show that a normalized index gets value infinity for every strictly positive cashflow and value zero if the cashflow is strictly negative:
$$
\alpha_t(D^{c,s})=+\infty\ \ \textrm{for $c>0$, and } \alpha_t(D^{c,s})=0\ \ \textrm{for $c<0$,}\ \ \textrm{for all } t\in \cT,
$$
where, for any $\omega \in \Omega$ and $s\geq t$, $D^{c,s}(r,\omega)=c$ for $r=s$ and zero otherwise.

For normalized DCAI we have equivalent forms of Property (D7).  In fact, one can show that under normalization, the set of properties (D1)--(D7) is equivalent to either the set (D1)--(D7-I) or the set (D1)--(D7-II), where
\begin{itemize}
\item[]{\bf (D7-I)} For a given $t\geq 0$ and $D,D'\in\mathcal{D}$, if $D_t(\omega)=D'_t(\omega)=0$ for all $\omega\in\Omega$, and there exists a non-negative $\mathcal{F}_t$-measurable random variable $m$ such that $\alpha_{t+1}(D,\omega)\geq m(\omega)\geq \alpha_{t+1}(D',\omega)$ for all $\omega\in\Omega$, then $\alpha_t(D,\omega)\geq m(\omega)\geq \alpha_t(D',\omega)$ for all $\omega\in\Omega$.
\item[]{\bf (D7-II)} For a given $t\geq0$ and $D\in\mathcal{D}$, if $D_t(\omega)=0$ for all $\omega\in\Omega$, then
    $$
    1_A\min_{\omega\in A}\alpha_{t+1}(D,\omega)\leq 1_A\alpha_t(D)\leq 1_A\max_{\omega\in A}\alpha_{t+1}(D,\omega)\,,
    $$
for all $A\in\mathcal{F}_t$.
\end{itemize}

\noindent
Finally we want to mention that (D3) and (D7) can be equivalently replaced in the definition of DCAI by the following two properties:
\begin{enumerate}
\item[]{\bf (D3-I)} For $D,D'\in\mathcal{D}$, if there exists $A\in\mathcal{F}_t$ such that $1_A D_s\geq 1_A D'_s$ for all $s\geq t$, then $1_A\alpha_t(D)\geq 1_A\alpha_t(D')$;
\item[]{\bf (D7-III)} For $D,D'\in\mathcal{D}$, if there exist $A\in\mathcal{F}_t$  and a
non-negative $\mathcal{F}_t$-measurable random variable $m$,  such that
$1_A D_t \geq 0 \geq 1_A D'_t$ and $1_A \alpha_{t+1}(D)\geq 1_A m\geq 1_A\alpha_{t+1}(D')$,
then $1_A \alpha_t(D)\geq 1_A m\geq 1_A \alpha_t(D')$.
\end{enumerate}

\section{Characterization of  Dynamic CAI by a family of Dynamic CRMs}\label{sec:DCRM}

As mentioned in Section \ref{sec:SAI}, there is a strong relationship between coherent acceptability indices and coherent risk measures. In fact, as seen from Theorem \ref{th-AI-static-Chernyi} and Theorem \ref{th:ReprThStaticCRM}, any CAI $\alpha$ can be represented in terms of a family of coherent risk measures $\rho^x, x\geq 0$:
\begin{equation}\label{eq:RepStaticCAIbyCRM}
\alpha(D)=\sup\{x\in[0,+\infty):\rho^x(D)\leq 0\}\,.
\end{equation}
Looking  at  \eqref{eq:RepStaticCAIbyCRM} one might think that a natural approach to constructing a DCAI would be to use
this representation but to replace the static coherent risk measures in \eqref{eq:RepStaticCAIbyCRM} by their dynamic counterpart. The representation \eqref{eq:AI-Representedby-RM} that we derive below shows that this is indeed the case. The delicate issue however is, what family of  dynamic coherent risk measures should be used. It turns out that in order to produce a DCAI satisfying a financially acceptable set of dynamic properties, one needs to use a carefully crafted family of dynamic coherent risk measures. In this section we introduce such a family of dynamic coherent risk measures and we compare our definition of coherent dynamic risk measures with analogous ones that have been already studied in the literature.

\subsection{Definition of dynamic coherent risk measure}

\begin{definition}
Dynamic coherent risk measure is a function $\rho: \{0,\ldots,T\}\times\mathcal{D}\times\Omega\to \mathbb{R}$ that satisfies the following properties:
\begin{enumerate}
\item[\bf (A1)] {\bf Adaptiveness.}
$\rho_t(D)$ is $\mathcal{F}_t$-measurable for all $t\in\cT$ and $D\in\mathcal{D}$;
\item[\bf (A2)] {\bf Independence of the past.}
If  $1_A D_s=1_A D'_s$ for some $t\in\cT$, $D, D'\in\mathcal{D}$, and $A\in\mathcal{F}_t$ and for all $s\geq t$, then $1_A\rho_t(D)=1_A\rho_t(D')$;
\item[\bf (A3)] {\bf Monotonicity.}
If $D_s(\omega)\geq D'_s(\omega)$ for some $t\in\cT$ and $D, D'\in\mathcal{D}$,   and for all $s\geq t$ and $\omega\in\Omega$,
then $\rho_t(D,\omega)\leq \rho_t(D',\omega)$ for all $\omega\in\Omega$;
\item[\bf (A4)] {\bf Homogeneity.}
$\rho_t(\lambda D,\omega) = \lambda \rho_t(D, \omega)$ for
all  $\lambda > 0, \ D\in\mathcal{D}, \ t\in\cT$, and $\omega\in\Omega$;
\item[\bf (A5)] {\bf Subadditivity.}
$\rho_t(D+D',\omega) \leq \rho_t(D,\omega) + \rho_t(D',\omega)$
for all $t\in\cT$, $D, D'\in\mathcal{D}$, and $\omega\in\Omega$;
\item[\bf (A6)] {\bf Translation invariance.}
$\rho_t(D+m1_{\{s\}})=\rho_t(D)-m$ for every
$t\in\cT$, $D\in\mathcal{D}$, $\mathcal{F}_t$-measurable random variable $m$, and all $s\geq t$;
\item[\bf (A7)] {\bf Dynamic consistency.}
$$
1_A (\min_{\omega\in A}\rho_{t+1}(D,\omega)-D_t) \leq 1_A \rho_t(D) \leq 1_A (\max_{\omega\in A}\rho_{t+1}(D,\omega)-D_t)\, ,
$$
for every $t\in\{0,1,\ldots, T-1\}$, $D\in\mathcal{D}$ and $A\in\mathcal{F}_t$.
\end{enumerate}
\end{definition}

\noindent We want to mention that our definition of DCRM differs from the definition given in previous studies essentially only by the
dynamic consistency property. For sake of completeness, we will present here how  property (A7) relates to other forms of dynamic
consistency of risk measures (for processes).
\begin{itemize}
\item[]{\bf (A7-I)} 
If $D_t=D'_t$, and $\rho_{t+1}(D)=\rho_{t+1}(D')$ for some
$t\in\{0,1,\ldots,T-1\}$, and $D,D'\in\mathcal{D}$,  then  $\rho_t(D)=\rho_t(D')$;
\item[]{\bf (A7-II)} 
$\rho_t(D)=\rho_t(-\rho_{t+1}(D)1_{\{t+1\}}) - D_t$ for
all times $t=0,1,\ldots,T-1$ and positions $D\in\mathcal{D}$.
\item[]{\bf(A7-III)} $\rho_t(D) \leq \rho_t(- \rho_{t+1}(D)1_{t+1}) - D_t$ for all $D\in\mathcal{D}, \ t\in\{0,1,\ldots,T-1\}$,
\item[]{\bf(A7-IV)} $\rho_t(D) \geq \rho_t(- \rho_{t+1}(D)1_{t+1}) - D_t$ for all $D\in\mathcal{D}, \ t\in\{0,1,\ldots,T-1\}$,
\item[]{\bf(A7-V)} if $D_t=0$, and $\rho_{t+1}(D) \leq 0$ for some $t\in\{0,1,\ldots,\}$ and $D\in\mathcal{D}$, then $\rho_t(D)\leq 0$.
\end{itemize}

 Property (A7-I) is the dynamic consistency property for DCRM defined by Riedel \cite{Riedel2004}.
Property (A7-II) is the version of the dynamic programming principle  (also called recursiveness),  introduced in Cheridito, Delbaen and Kupper  \cite{CheriditoDelbaenKupper2006}, adapted to the  setup of our paper, that is, it is stated in terms of dividend processes rather than value process as in \cite{CheriditoDelbaenKupper2006}. Properties (A7-I) and (A7-II) are equivalent, and they are also sometimes called {\it strong dynamic consistency property}. To the best of our knowledge, properties (A7-III) and (A7-IV) were first introduced in the context of random processes in Acciaio, F\"{o}llmer and Penner~\cite{AcciaioFollmerPenner2010}, and they were called {\it acceptance and rejection consistency}, respectively. In the same paper, Acciaio, F\"{o}llmer and Penner introduced condition (A7-V) and they called it \textit{weakly acceptance consistent}.

For corresponding definitions in case of random variables rather than random processes we refer to the survey paper by Acciaio and Penner \cite{AcciaioPenner2010} and references therein.

It easy to show that the dynamic consistency condition (A7) is stronger than (A7-V), and it is weaker than (A7-I) or (A7-II). Also note that since conditions (A7-II) and (A7-III) taken together are equivalent to (A7-II), then, taken together they imply (A7). However, the inverse implication is not necessarily true.

We conclude this subsection with  the following result.

\begin{proposition}
If $\rho$ is a dynamic coherent risk measure, then
$\rho_t(c1_{\{s\}},\omega)=-c$, for all $c\in\mathbb{R}$, $t\in\cT$, $\omega\in\Omega$ and $s\geq t$.
\end{proposition}
\proof Given some fixed $t\in\cT$ and $\omega\in\Omega$, denote by
$\lambda:=\rho_t(0,\omega)$. Then, by translation invariance (A6) of $\rho$, we deduce
\begin{equation}\label{eq:normalityRho}
\rho_t(c1_{\{s\}},\omega) =\rho_t(0,\omega)-c = \lambda-c \, ,
\end{equation}
for all $c\in\mathbb{R}$.  In particular, for $c=1$, we have
$\rho_t(1_{\{s\}},\omega)=\lambda-1$.
Hence, by (A4)-homogeneity of $\rho$, it follows that
$\rho_t(c_u1_{\{s\}},\omega)=c_u\rho_t(1_{\{s\}},\omega) =c_u(\lambda-1)$, for all $c_u>0$.
Combining this with \eqref{eq:normalityRho} we get
$\lambda-c_u=c_u\lambda-c_u$, and consequently
$\lambda(1-c_u)=0$, for arbitrary positive $c_u$. Thus, we conclude that $\lambda=0$, and hence
$\rho_t(0,\omega)=0$. With this, by \eqref{eq:normalityRho}, the proposition follows.
\endproof
Note that, in particular, $\rho_t(0)=0$, for all $t\in\cT$.

\subsection{Duality between DCAI and DCRM}\label{subsec:dualityDCAIvsDCRM}
We start this section with several definitions that will be used in the main results derived here.

\begin{definition}
A family of dynamic coherent risk measures $(\rho^x)_{x\in(0,+\infty)}$ is called {\it increasing}
if $\rho^x_t(D,\omega) \geq \rho^y_t(D,\omega)$, for all $x\geq y > 0$, $t\in\cT$, $D\in\mathcal{D}$ and $\omega\in\Omega$.
\end{definition}

\begin{definition}
A dynamic acceptability index $\alpha$ is called {\it right-continuous}
if $\lim\limits_{c\to 0^+}\alpha_t(D+c1_{\{t\}},\omega)=\alpha_t(D,\omega)$,
for all $t\in\cT$, $D\in\mathcal{D}$, and $\omega\in\Omega$.
\end{definition}

\begin{definition}
A family of dynamic coherent risk measures $(\rho^x)_{x\in(0,+\infty)}$ is called {\it left-continuous} at $x_0>0$,
if $\lim\limits_{x\to x_0^-}\rho^x_t(D,\omega)=\rho^{x_0}_t(D,\omega)$, for all $t\in\cT$, $D\in\mathcal{D}$, and $\omega\in\Omega$.
\end{definition}

\begin{theorem}\label{th:RM-Representedby-AI}
Assume that $\alpha$ is a normalized dynamic coherent acceptability index. Then, the set of functions
$\rho^x, x\in\mathbb{R}$, defined by
\begin{equation}\label{eq:RM-Representedby-AI}
\rho^x_t(D,\omega):=\inf\{c \in\mathbb{R}:\alpha_t(D+c1_{\{t\}},\omega)\geq x\}\,,
\end{equation}
for all $t\in\cT$, $D\in\mathcal{D}$ and $\omega\in\Omega$,
is an increasing, left-continuous family of dynamic coherent risk measures.
\end{theorem}
\proof
First we will show that $\rho^x$ defined by \eqref{eq:RM-Representedby-AI} is well-defined.
Since $\alpha$ is normalized, for all $t\in\cT$, $D\in\mathcal{D}$, there exist two finite constants $c^{t,D}_u$ and $c^{t,D}_l$ such that
$$
\alpha_t(D+c^{t,D}_u1_{\{t\}},\omega)=+\infty \textrm { and } \alpha_t(D+c^{t,D}_l1_{\{t\}},\omega)=0\,,
$$
for all $\omega\in\Omega$. Hence, for every $x\in(0,+\infty)$, the set $\{c \in\mathbb{R}:\alpha_t(D+c1_{\{t\}},\omega)\geq x\}$ is not empty,
and
$c^{t,D}_l\leq \inf\{c \in\mathbb{R}:\alpha_t(D+c1_{\{t\}},\omega)\geq x\}$.
From here we conclude that infimum from \eqref{eq:RM-Representedby-AI} is finite, and hence $\rho^x$ is well-defined.

Next we will show that $\rho^x, x\in(0,+\infty)$, satisfies properties (A1)-(A7).
By (D1)-adapt\-iveness and (D2)-independence of the past of $\alpha$, property (A1) and (A2) for $\rho^x, \ x\in\mathbb{R}$, follow immediately.

Assume that $t\in\cT$ and $D, D'\in\mathcal{D}$ are such that
$D_s(\omega)\geq D'_s(\omega)$ for all $s\geq t$ and $\omega\in\Omega$.
Then
$(D+c1_{\{t\}})_s(\omega)\geq (D'+c1_{\{t\}})_s(\omega)$ for $s\geq t, \ \omega\in\Omega$,  and $c\in\mathbb{R},$
and by (D3), monotonicity of $\alpha$
\begin{align}\label{proof:AI_RM_DUALITY3}
\alpha_t(D+c1_{\{t\}},\omega)\geq\alpha_t(D'+c1_{\{t\}},\omega)\, ,
\end{align}
for all $c\in\mathbb{R}$ and $\omega\in\Omega$. From here, we deduce the following inclusion
$$
\{c \in\mathbb{R}:\alpha_t(D+c1_{\{t\}},\omega)\geq x\}\supseteq \{c \in\mathbb{R}:\alpha_t(D'+c1_{\{t\}},\omega)\geq x\}\,.
$$
Taking infimum of both sets, $(A3)$ follows. Similarly, the homogeneity of $\rho^x$ follows from the scale invariance of $\alpha$.

Next we show that $\rho^x$ satisfies $(A5)$. Let $t\in\cT$, $D, D'\in\mathcal{D}$ and
$\omega\in\Omega$, and let us take $c_1, c_2\in \mathbb{R}$ such that
$$
\alpha_t(D+c_1 1_{\{t\}},\omega)\geq x\,, \quad \alpha_t(D'+c_2 1_{\{t\}},\omega)\geq x \, .
$$
Then, by (D5), quasi-concavity of $\alpha$, we have
\begin{align*}
\alpha_t({1\over2}D+{1\over2}c_11_{\{t\}}+{1\over2}D'+{1\over2}c_21_{\{t\}},\omega)\geq x\,,
\end{align*}
and therefore by (D4), scale invariance of $\alpha$, we get
$\alpha_t(D+D'+(c_1+c_2)1_{\{t\}},\omega) \geq x$.
 This implies that $c_1+c_2\in \{c \in\mathbb{R}: \alpha_t(D+D'+c1_{\{t\}},\omega)\geq x\}$.
Hence,
\begin{align}
c_1+c_2&\geq \inf\{c\in\mathbb{R}: \alpha_t(D+D'+c1_{\{t\}},\omega)
\geq x\}\nonumber\\
&=\rho^x_t(D+D',\omega) \label{eq:AI_RM_DUALITY5}\,.
\end{align}
 Note that the above inequality holds true for all $c_1\in\{c \in\mathbb{R}: \alpha_t(D+c1_{\{t\}},\omega)\geq x\}$ and $c_2\in\{c \in\mathbb{R}: \alpha_t(D'+c1_{\{t\}},\omega)\geq x\}$. By taking infimum in \eqref{eq:AI_RM_DUALITY5},
first with respect to $c_1$, and then with respect to $c_2$, we have,
$\rho^x_t(D,\omega)+\rho^x_t(D',\omega)\geq\rho^x_t(D+D',\omega)$, and hence $(A5)$ is checked.

Now we will show that $\rho^x$ satisfies (A6), translation invariance.
Fix  an $\omega^0\in\Omega$, $t\in\cT$, $D\in\mathcal{D}$ and an $\mathcal{F}_t$-measurable random variable $m$.
Denote by $P^t_i$ the unique element  of partition of $\mathcal{F}_t$ such that  $\omega^0\in P^t_i$.
This yields that the cash-flows $m1_{\{t\}}$ and $m(\omega^0)1_{\{t\}}$ agree on the set $P^t_i$, and for all times $s\geq t$.
Then, for any constant $c\in\mathbb{R}$, we have
$$
1_{P^t_i}(D+m1_{t}+c1_{\{t\}})_s=1_{P^t_i}(D+
m(\omega^0)1_{\{t\}}+c1_{\{t\}})_s\,, \textrm{ for } s\geq t\,.
$$
By (D2), independence of the past of $\alpha$, we have
$$
1_{P^t_i}\alpha_t(D+m1_{t}+c1_{\{t\}})=1_{P^t_i}\alpha_t(D
+m(\omega^0)1_{\{t\}}+c1_{\{t\}})\,.
$$
Since $m$ is $\mathcal{F}_t$-measurable, by (D6), translation invariance of $\alpha$, we have
$$
\alpha_t(D+m1_{s}+c1_{\{t\}},\omega^0)=\alpha_t(D+m1_{t}+c1_{\{t\}},\omega^0)\,, \quad \textrm{for all } s\geq t.
$$
Combining the above with \eqref{eq:RM-Representedby-AI}, we deduce
\begin{align*}
\rho^x_t(D+m1_{\{s\}},\omega^0)
&=\inf\{c\in\mathbb{R}: \alpha_t(D+m1_{\{s\}}+c1_{\{t\}},\omega^0)\geq x\}\\
&=\inf\{c \in\mathbb{R}: \alpha_t(D+m1_{\{t\}}+c1_{\{t\}},\omega^0)\geq x\}\\
&=\inf\{c \in\mathbb{R}: \alpha_t(D+m(\omega^0)1_{\{t\}}+c1_{\{t\}},\omega^0)\geq x\}\\
&=\inf\{m(\omega^0)+c \in\mathbb{R}: \alpha_t(D+(m(\omega^0)+c)1_{\{t\}},\omega^0)\geq x\}-m(\omega^0)\\
&=\rho^x_t(D,\omega)-m(\omega^0)\,.
\end{align*}
Since $\omega^0$ is arbitrarily chosen in $\Omega$, we obtain $\rho^x_t(D+m1_{\{s\}})=\rho^x_t(D)-m$,
for all $s\geq t$, and (A6) is checked.

Next we will show that $\rho^x$ satisfies $(A7)$, dynamic consistency.
Assume that $t\in\cT$, $D\in\mathcal{D}$ and $A\in\mathcal{F}_t$ are fixed, and denote by
$c^{t,D,A}_{\min}:=\min\limits_{\omega\in A}\rho^x_{t+1}(D,\omega)$ and
$c^{t,D,A}_{\max}:=\max\limits_{\omega\in A}\rho^x_{t+1}(D,\omega)$.
Then $\alpha_{t+1}(D+c_01_{\{t+1\}},\omega)<x$, for all $\omega\in A$ and for any $c_0<c^{t,D,A}_{\min}$.
Due to the finiteness of the probability space $\Omega$, there exists a number $\epsilon_{A} >0$, such that
$\alpha_{t+1}(D+c_01_{\{t+1\}},\omega)\leq x-\epsilon_{A}$,
for all $\omega\in A$. By (D2), independent of the past of $\alpha$, we have
$$
\alpha_{t+1}(D-D_t1_{\{t\}}+c_01_{\{t+1\}},\omega)=\alpha_{t+1}(D+c_01_{\{t+1\}},\omega)\leq x-\epsilon_{A}\,,
$$
for all $\omega\in A$. Since, $1_A(D-D_t1_{\{t\}}+c_01_{\{t+1\}})_t=1_A(D_t-D_t)=0$, then,  by (D7)
$$
\alpha_{t}(D-D_t1_{\{t\}}+c_01_{\{t+1\}},\omega)\leq x-\epsilon_{A}, \quad \omega\in A\, .
$$
Consequently, since $c_0$ is a constant, by (D6)
\begin{align*}
\alpha_{t}(D+(c_0-D_t)1_{\{t\}},\omega)&=\alpha_{t}(D-D_t1_{\{t\}}+c_01_{\{t\}},\omega)\\
&=\alpha_{t}(D-D_t1_{\{t\}}+c_01_{\{t+1\}},\omega)\\
&\leq x-\epsilon_{A}<x\, ,
\end{align*}
for all $\omega\in A$ and $c_0<c^{t,D,A}_{\min}$.
By the definition of $\rho^x$, we get
$$
\rho^x_t(D,\omega)=\inf\{c\in\mathbb{R}:\alpha_t(D+c1_{\{t\}},\omega)\geq x\}\geq c_0-D_t(\omega)\, ,
$$
for all $\omega\in A$ and $c_0<c^{t,D,A}_{\min}$.
Hence, $\rho^x_t(D,\omega)\geq c^{t,D,A}_{\min}-D_t(\omega)$, or equivalently
$1_A\rho^x_t(D)\geq 1_A(\min_{\omega\in A}\rho^x_{t+1}(D,\omega)-D_t)$.
Similarly, one can show that
$$
1_A\rho^x_t(D)\leq 1_A(\max_{\omega\in A}\rho_{t+1}(D,\omega)-D_t),
$$
and thus (A7) is established.

All the above imply that $\rho^x$ is a DCRM for every $x>0$.

Monotonicity of $\rho^x$ with respect to $x$ follows immediately from \eqref{eq:RM-Representedby-AI} and the inclusion
$$
\{c\in\mathbb{R}:\alpha_t(D+c1_{\{t\}},\omega)\geq x\}\subseteq\{c\in\mathbb{R}:\alpha_t(D+c1_{\{t\}},\omega)\geq y\}, \quad x\geq y>0\, .
$$

Finally, we will show that $(\rho^x)_{x\in(0,+\infty)}$ is left-continuous.
Let $x_0$ be any positive number. Then,
\begin{equation}\label{eq:proof_AI_RM_duality15}
\inf\{c \in\mathbb{R}:\alpha_t(D+c1_{\{t\}},\omega)\geq x_0\}\geq \lim_{x\to x^-_0}\inf\{c \in\mathbb{R}:\alpha_t(D+c1_{\{t\}},\omega)\geq x\}\,.
\end{equation}
If the above inequality holds strictly, then there exists a constant $c_0$ such that,
\begin{align}\label{proof:AI_RM_DUALITY14}
\inf\{c \in\mathbb{R}:\alpha_t(D+c1_{\{t\}},\omega)\geq x_0\}>c_0>\lim_{x\to x^-_0}\inf\{c \in\mathbb{R}:\alpha_t(D+c1_{\{t\}},\omega)\geq x\}\,.
\end{align}
Note that $\inf\{c \in\mathbb{R}:\alpha_t(D+c1_{\{t\}},\omega)\geq x\}$ is an non-decreasing function with respect to $x$.
Therefore, the second inequality in \eqref{proof:AI_RM_DUALITY14} implies that,
$c_0>\inf\{c \in\mathbb{R}:\alpha_t(D+c1_{\{t\}},\omega)\geq x\}$,
for all $x<x_0$. Hence, by (D3), monotonicity of $\alpha$,
$\alpha_t(D+c_01_{\{t\}},\omega)\geq x$, for all $x<x_0$, and thus
$\alpha_t(D+c_01_{\{t\}},\omega)\geq \lim_{x\to x^-_0} x=x_0$.
On the other hand, by the first inequality in \eqref{proof:AI_RM_DUALITY14}, we deduce that,
$\alpha_t(D+c_01_{\{t\}},\omega)<x_0.$
Contradiction. Therefore, we should have strict equality in \eqref{eq:proof_AI_RM_duality15}.

This completes the proof.
\endproof

\noindent
Next Theorem shows the representation of a DCAI in terms of a family of DCRMs.

\begin{theorem}\label{th:AI-Representedby-RM}
Assume that $(\rho^x)_{x\in(0,+\infty)}$ is an increasing family of dynamic coherent risk measures.
Then the function $\alpha$ defined as follows,
\begin{equation}\label{eq:AI-Representedby-RM}
\alpha_t(D,\omega) :=\sup\{x\in(0,+\infty): \rho^x_t(D,\omega)\leq 0\}\,,
\end{equation}
for $t\in\cT$, $D\in\mathcal{D}$ and $\omega\in\Omega$, is a normalized, right-continuous, dynamic coherent acceptability index.
Here, we assume $\sup\emptyset=0$.
\end{theorem}
\proof
Note that the assumption $\sup\emptyset=0$ guarantees that $\alpha$ from
\eqref{eq:AI-Representedby-RM} is well-defined and takes values in $[0,+\infty]$.

In the following, we will prove that $\alpha$ defined in \eqref{eq:AI-Representedby-RM}
satisfies properties (D1)--(D7).

(D1) - adaptiveness, (D2) - independence of the past, (D4) - scale invariance, and (D6) - translation invariance  follow immediately from the definition of $\alpha$,  and from
adaptiveness (A1), independence of the past (A2), homogeneity (A4)  and translation invariance (A6) of $\rho^x$, respectively.

Let $t\in\cT$, $D,D'\in\mathcal{D}$,
and assume that $D_s(\omega)\geq D'_s(\omega)$ for all $s\geq t$, and $\omega\in\Omega$.
By (A3), monotonicity of $\rho^x$, we have
\begin{align}\label{proof:AI_RM_DUALITY6}
\rho^x_t(D)\leq \rho^x_t(D')\,, \quad \textrm{ for all }  x>0 \, .
\end{align}
Note that, for any $x_0\in\{x\in(0,+\infty): \rho^x_t(D',\omega)\leq 0\}$, we have
$\rho^{x_0}_t(D',\omega)\leq 0$, which combined with \eqref{proof:AI_RM_DUALITY6} implies
$\rho^{x_0}_t(D,\omega)\leq \rho^{x_0}_t(D',\omega)\leq 0\, , \quad \omega\in\Omega$.
Therefore,
$$
\{x\in(0,+\infty): \rho^x_t(D,\omega)\leq 0\}\supseteq \{x\in(0,+\infty): \rho^x_t(D',\omega)\leq 0\}
$$
By taking supremum of both sides, (D3) follows.

Next we will prove that $\alpha$ is quasi-concave.
For given $t\in\cT$, and $x^0\in(0,+\infty]$, if
$D,D'\in\mathcal{D}$ are such that $\alpha_t(D,\omega)\geq x^0, \alpha_t(D',\omega)\geq x^0$,
then, by definition \eqref{eq:AI-Representedby-RM} of $\alpha$,
and monotonicity of $\rho^x$ in $x$, we conclude that for any $x<x^0$,
\begin{align*}
\rho^x_t(D,\omega)\leq 0\,, \quad
\rho^x_t(D',\omega)\leq 0\,.
\end{align*}
By (A4), homogeneity of $\rho^x$, we note that for any $\lambda \in[0,1]$ and $x<x^0$,
\begin{align*}
\rho^x_t(\lambda D,\omega)=\lambda\rho^x_t(D,\omega) \leq 0\, , \quad
\rho^x_t((1-\lambda)D',\omega)=(1-\lambda)\rho^x_t(D',\omega)\leq 0\,.
\end{align*}
From here, by (A5), subadditivity of $\rho^x$, we get
$$
\rho^x_t(\lambda D+(1-\lambda)D',\omega)\leq \rho^x_t(\lambda D,\omega)+\rho^x_t((1-\lambda)D',\omega)\leq 0\, ,
$$
for any $x<x^0$.
Hence
$\sup\{x\in(0,+\infty): \rho^x_t(\lambda D+(1-\lambda)D',\omega)\leq 0\}\geq x^0$, and thus,
by definition \eqref{eq:AI-Representedby-RM} of $\alpha$, we have,
$\alpha(\lambda D+(1-\lambda)D',\omega) \geq x_0$. This yields quasi-concavity of $\alpha$.

Assume that $D\in\mathcal{D}$, and $m$ is an $\mathcal{F}_t$-measurable random variable.
By \eqref{eq:AI-Representedby-RM} and (A6), we get
\begin{align*}
\alpha_t(D+m1_{\{s\}},\omega)&=\sup\{x\in(0,+\infty): \rho^x_t(D+m1_{\{s\}},\omega)\leq 0\} \\
&=\sup\{x\in(0,+\infty): \rho^x_t(D+m1_{\{t\}},\omega)\leq 0\}\\
&=\alpha_t(D+m1_{\{t\}},\omega)\, ,
\end{align*}
for all $s\geq t$ and $\omega\in\Omega$. Hence, $\alpha$ satisfies property (A6).

Now, let us show that $\alpha$ satisfies dynamic consistency property (D7).
Assume that $D,D'\in\mathcal{D}$, and $t\in\cT$ are such that
$D_t(\omega) \geq 0 \geq D'_t(\omega)$ for all $\omega\in\Omega$, and there exists a non-negative
$\mathcal{F}_t$-measurable random variable $m$ such that $\alpha_{t+1}(D,\omega)\geq m(\omega) \geq \alpha_{t+1}(D',\omega)$ for all $\omega\in\Omega$.
By definition \eqref{eq:AI-Representedby-RM}
$$
\sup\{x\in(0,+\infty): \rho^x_{t+1}(D,\omega)\leq 0\} \geq m(\omega) \geq \sup\{x\in(0,+\infty): \rho^x_{t+1}(D',\omega)\leq 0\}\, ,
$$
for all $\omega\in\Omega$. Let us fix an $\bar{\omega}\in \Omega$, and denote by $\bar{c}:=m(\bar{\omega})$.
There exists a $P^t_i$ such that $\bar{\omega}\in P^t_i$. From the above inequality, we conclude that for all $\omega\in P^t_i$
$$
\sup\{x\in(0,+\infty): \rho^x_{t+1}(D,\omega)\leq 0\} \geq \bar{c} \geq \sup\{x\in(0,+\infty): \rho^x_{t+1}(D',\omega)\leq 0\}\,.
$$
Then, for all $c'> \bar{c}$ and $\omega\in P^t_i$, $c'>\sup\{x\in(0,+\infty): \rho^x_{t+1}(D',\omega)\leq 0\}$,
which consequently implies that
\begin{align}\label{proof:AI_RM_DUALITY7}
\rho^{c'}_{t+1}(D',\omega)>0\,.
\end{align}
Also note that $\sup\{x\in(0,+\infty): \rho^x_{t+1}(D,\omega)\leq 0\}>c$, for any $c<\bar{c}$.
By monotonicity of $\rho^x$ with respect to $x$, we have
$\rho^{c}_{t+1}(D,\omega)\leq 0, \ \omega\in P_i^t$.
Due to the finiteness of $\Omega$, \eqref{proof:AI_RM_DUALITY7} implies that $\min_{\omega\in P^t_i}\rho^{c'}_{t+1}(D',\omega)>0$,
for all $c'>\bar{c}$.
Using (A7), dynamic consistency of $\rho^x$, we get the following
\begin{align*}
1_{P^t_i}\rho^{c'}_t(D')&\geq 1_{P^t_i}(\min_{\omega\in P^t_i}\rho^{c'}_{t+1}(D',\omega)-D'_t)\\
&=1_{P^t_i}\min_{\omega\in P^t_i}\rho^{c'}_{t+1}(D',\omega)-1_{P^t_i}D'_t,  \quad c'>\bar{c} \,.
\end{align*}
Equivalently,
\begin{align}\label{proof:AI_RM_DUALITY8}
\rho^{c'}_t(D',\omega)=\min_{\omega\in P^t_i}\rho^{c'}_{t+1}(D',\omega)-D'_t(\omega)>-D'_t(\omega)\geq 0\, ,
\end{align}
for all $\omega\in P^t_i$, and $c'>\bar{c}$.

If $\bar{c} < \sup\{x\in(0,+\infty): \rho^x_{t}(D',\omega')\leq 0\}$,
for some  $\omega'\in P^t_i$, then there exists a constant $c^0$ such that
$$
\bar{c} < c^0 < \sup\{x\in(0,+\infty): \rho^x_{t}(D',\omega')\leq 0\}\,.
$$
This implies that $\rho^{c^0}_{t}(D',\omega')\leq 0$, that contradicts
with \eqref{proof:AI_RM_DUALITY8}. Therefore,
$$
\bar{c}\geq \sup\{x\in(0,+\infty): \rho^x_{t}(D',\omega)\leq 0\}\, ,
$$
and by \eqref{eq:AI-Representedby-RM}, we have
\begin{equation}\label{eq:AI-RepTh-D6-1}
\bar{c}\geq \alpha_t(D',\omega)\,, \quad \omega\in P^t_i.
\end{equation}
By similar arguments, one can show that
\begin{equation}\label{eq:AI-RepTh-D6-2}
\bar{c}\leq \alpha_t(D,\omega)\,, \quad \omega\in P^t_i.
\end{equation}
Since $\bar{\omega}$ was arbitrarily chosen, by \eqref{eq:AI-RepTh-D6-1} and \eqref{eq:AI-RepTh-D6-2}, we finally conclude that,
$$
\alpha_t(D,\omega)\geq m(\omega)\geq \alpha_t(D',\omega), \quad \textrm{for all } \ \omega\in\Omega\,.
$$
Thus (A7) is checked.

Let us show that $\alpha$ is right-continuous. Given $t\in\cT, D\in \mathcal{D}$ and $\omega\in\Omega$, we have
$$
\{x\in(0,+\infty): \rho^x_t(D,\omega)\leq 0\}\subseteq \{x\in(0,+\infty): \rho^x_t(D,\omega)\leq c\}\, ,
$$
for any constant $c>0$. Taking the supremum of both sides, and then the limit of the right hand side as $c\to 0+$,
we get
\begin{equation} \label{eq:AI-RepTh-Cont0}
\sup\{x\in(0,+\infty): \rho^x_t(D,\omega)\leq 0\}\leq\lim_{c\to 0^+}\sup\{x\in(0,+\infty): \rho^x_t(D,\omega)\leq c\}\,.
\end{equation}
If the above inequality holds strictly, then  there exists $x^0\in(0,+\infty)$ such that
\begin{equation}\label{eq:AI-RepTh-Cont1}
\sup\{x\in(0,+\infty): \rho^x_t(D,\omega)\leq 0\}< x^0 < \lim_{c\to 0^+}\sup\{x\in(0,+\infty): \rho^x_t(D,\omega)\leq c\}\,.
\end{equation}
The second inequality implies that
$$
x^0<\sup\{x\in(0,+\infty): \rho^x_t(D,\omega)\leq c\}, \quad \textrm{for all} \  c>0.
$$
By monotonicity of $\rho^x$, we deduce that $\rho^{x^0}_t(D,\omega)\leq c$.
Since the last inequality holds true for all $c>0$, we have that
$\rho^{x^0}_t(D,\omega)\leq\lim_{c\to0^+}c=0$,
that contradicts with first strict inequality in \eqref{eq:AI-RepTh-Cont1}.
Therefore, we have equality in \eqref{eq:AI-RepTh-Cont0}. Using this equality,  and
(A6), translation invariance of $\rho^x$, we write
\begin{align*}
\alpha_t(D, \omega) &= \sup\{x\in(0,+\infty): \rho^x_t(D,\omega)\leq 0\}\\
&=\lim_{c\to 0^+}\sup\{x\in(0,+\infty): \rho^x_t(D,\omega)\leq c\}\\
&=\lim_{c\to 0^+}\sup\{x\in(0,+\infty): \rho^x_t(D+c1_{\{t\}},\omega)\leq 0\}\\
&=\lim_{c\to 0^+}\alpha_t(D+c1_{\{t\}},\omega)\, ,
\end{align*}
and right continuity of $\alpha$ is established.

Finally, we will prove that $\alpha$ is normalized. Given a fixed $t\in\cT$, consider the following cash-positions
\begin{equation*}
D_{pos} :=1_{\{t\}}, \quad D_{neg}:=-1_{\{t\}}\,.
\end{equation*}
Recall that $\rho_t(0)=0$. By \eqref{eq:AI-Representedby-RM} and (A6), we have
\begin{align*}
\alpha_t(D_{pos},\omega)
&=\sup\{x\in(0,+\infty): \rho^x_t(1_{\{t\}},\omega)\leq 0\}\\
&=\sup\{x\in(0,+\infty): \rho^x_t(0,\omega)-1\leq 0\}\\
&=\sup\{x\in(0,+\infty): -1\leq 0\} = +\infty\,.
\end{align*}
Similarly, one can show that $\alpha_t(D_{neg},\omega)=0$. \\
The proof is complete.
\endproof

We conclude this section with the main  result that provides a representation of a DCAI in terms of a family of DCRMs, and vise versa,  a representation of DCRM in terms of a DCAI.

\begin{theorem} \mbox{} \label{th:AI-RM-DUALITY}
\begin{itemize}
\item[(i)]
If $\alpha$ is a normalized, right-continuous, dynamic coherent acceptability index,
then there exists a left-continuous and increasing family of dynamic coherent risk measures \\ $(\rho^x)_{x\in(0,+\infty)}$, such that
\begin{equation}
\alpha_t(D,\omega)=\sup\{x\in(0,+\infty): \rho^x_t(D,\omega)\leq 0\}\, .
\end{equation}
\item[(ii)] If $(\rho^x)_{x\in(0,+\infty)}$ is a left-continuous and increasing family of dynamic coherent risk measures,
 then there exists a right-continuous and normalized dynamic coherent acceptability index $\alpha$ such that,
\begin{align*}
\rho^x_t(D,\omega):=\inf\{c \in\mathbb{R}:\alpha_t(D+c1_{\{t\}},\omega)\geq x\}\,,
\end{align*}
\end{itemize}
Here we assume that $\inf\emptyset=\infty$ and $\sup\emptyset=0$.
\end{theorem}

\proof (i)
For every $x\in(0,+\infty)$, define $\rho^x=(\rho^x_t)_{t=0}^T$ as follows,
\begin{equation}\label{proof:AI_RM_DUALITY12}
\rho^x_t(D,\omega):=\inf\{c \in\mathbb{R}:\alpha_t(D+c1_{\{t\}},\omega)\geq x\}\,,
\end{equation}
for all $t\in\cT$, $D\in\mathcal{D}$ and $\omega\in\Omega$.
By theorem \ref{th:RM-Representedby-AI}, $(\rho^x)_{x\in(0,+\infty)}$ is an increasing, left-continuous, family of  dynamic coherent risk measures.
We will show that
$$
\alpha_t(D,\omega)=\sup\{x\in(0,+\infty): \rho^x_t(D,\omega)\leq 0\}\, ,
$$
for all $t\in\cT, D\in \mathcal{D}$ and $\omega\in\Omega$.

For any $t\in\cT, D\in \mathcal{D}$, $\omega\in\Omega$, and $y^{t,D,\omega}>\sup\{x\in(0,+\infty): \rho^x_t(D,\omega)\leq 0\}$, we have
$$
\rho^{y^{t,D,\omega}}_t(D,\omega)>0\,.
$$
By \eqref{proof:AI_RM_DUALITY12}
$\inf\{c\in\mathbb{R}:\alpha_t(D+c1_{\{t\}},\omega)\geq y^{t,D,\omega}\}>0$,
and hence,
$$
\alpha_t(D,\omega)=\alpha_t(D+01_{\{t\}},\omega)<y^{t,D,\omega}\,.
$$
Since the above inequality holds true for all $y^{t,D,\omega}>\sup\{x\in(0,+\infty): \rho^x_t(D,\omega)\leq 0\}$, we conclude
that
$$
\alpha_t(D,\omega)\leq \sup\{x\in(0,+\infty): \rho^x_t(D,\omega)\leq 0\}\,.
$$
Similarly, one can show that $\alpha_t(D,\omega)\geq \sup\{x\in(0,+\infty): \rho^x_t(D,\omega)\leq 0\}$.

\noindent (ii) Define the function $\alpha$ as follows,
\begin{align}\label{proof:AI_RM_DUALITY15}
\alpha_t(D,\omega):=\sup\{x\in(0,+\infty): \rho^x_t(D,\omega)\leq 0\}\, ,
\end{align}
for all $t\in\cT$, $D\in\mathcal{D}$ and $\omega\in\Omega$.
By theorem \eqref{th:AI-Representedby-RM}, $\alpha$ is a right-continuous and normalized dynamic coherent acceptability index.
Finally, one can check that
$$
\rho^x_t(D,\omega):=\inf\{c\in\mathbb{R}:\alpha_t(D+c1_{\{t\}},\omega)\geq x\}\,,
$$
for all $x\in(0,+\infty)$, $t\in\cT, D\in \mathcal{D}$ and $\omega\in\Omega$.
\endproof

\section{Special Construction of DCAIs}\label{sec:SCCDAI}

It is known, that a dynamic coherent risk measure has a representation similar to \eqref{eq:repTh-staticRM}.
One of the important discoveries done in the process of robust representation of dynamic risk measures, similar to  \eqref{eq:repTh-staticRM}, was that due to dynamic consistency property (A7), the set of probability measures $\cD$ has to posses some additional features, which depend on how the dynamic consistency property (A7) is formulated.
A set of probability measures having such additional features is referred to as a dynamically consistent set of probability measures (or, for short, a consistent set of probability measures).

In Section \ref{subsec:ConsistentSetofPM} we present our version of the  concept of dynamically consistent set of probability measures, as well as some non-trivial examples of such sets. It is seen that our concept is different from the ones previously studied in the literature. Its form and properties have been dictated by the goal of using it in the context of robust representation of our DCAI.

 In Section \ref{subsec:reprThforDCRM} we prove the representation theorem for DCRM in terms of  consistent sets of probability measures. We conclude this section with representation theorem for DCAIs in terms of families of sequences of dynamically  consistent sets of probability measures.

\subsection{Dynamically consistent sequence of sets of probability measures}\label{subsec:ConsistentSetofPM}

In this section we shall discuss the concept of dynamically consistent sequence of sets of probability measures.

In what follows we denote by $\mathcal{P}$ the set of all absolutely continuous probability measures with respect to the underlying probability $\mathbb{P}$,
and $\mathcal{P}^e$ will stand for the set of all equivalent probability measures with respect to $\mathbb{P}$. Recall that our standing assumption is that $\mathbb{P}$ has full support. Note that in this case, due to the finiteness  of $\Omega$, the set $\cP$ consists of all probability measures on $\Omega$, and also
$\cP^e$ coincides with the set of all probability measures on $\Omega$ of full support.
\begin{definition}\label{def:ConsistentSetProbability}
A sequence of sets of probability measures $\{\mathcal{Q}_t\}^T_{t=0}$, with $\mathcal{Q}_t\subseteq\mathcal{P}$, is called
{\it dynamically consistent} with respect to the filtration $\mathbb{F}$, if the following inequalities hold true
\begin{equation}\label{eq:dynamic-cons}
1_A\min_{\omega\in A}\bigg\{\inf_{\mathbb{Q}\in\mathcal{Q}_{t+1}}\mathbb{E}_{\mathbb{Q}}[X|\mathcal{F}_{t+1}](\omega)\bigg\}\leq 1_A\inf_{\mathbb{Q}\in\mathcal{Q}_{t}}\mathbb{E}_{\mathbb{Q}}[X|\mathcal{F}_{t}]\leq 1_A\max_{\omega\in A}\bigg\{\inf_{\mathbb{Q}\in\mathcal{Q}_{t+1}}\mathbb{E}_{\mathbb{Q}}[X|\mathcal{F}_{t+1}](\omega)\bigg\}\,,
\end{equation}
for every $t\in\{0,\dots,T-1\}$, $A\in\mathcal{F}_t$, and every random variable $X$.
\end{definition}

\begin{definition}\label{def:StrongConsistentSetProbability}
A set of probability measures $\mathcal{Q}\subseteq\mathcal{P}$ is called
{\it consistent} with respect to filtration $\mathbb{F}$, if
the following equality holds true
\begin{equation}\label{eq:set-consistent}
\inf\limits_{Q\in\mathcal{Q}} \mathbb{E}_{Q}\Big[ X \, | \, \mathcal{F}_t \Big] =
\inf\limits_{Q\in\mathcal{Q}} \mathbb{E}_Q\Big[
\inf\limits_{M\in\mathcal{Q}} \mathbb{E}_M \big[   X  \, |\, \mathcal{F}_{t+1} \big] \, | \, \mathcal{F}_t \Big] \, ,
\end{equation}
for every $t\in\{0,\dots,T-1\}$, and every random variable $X$.
\end{definition}

\bigskip \noindent

\begin{proposition}\label{prop:cons2DCons}
If a set of probability measures $\cQ\subseteq\mathcal{P}$ is consistent, then $\{\cQ_t\}_{t=0}^T$ with $ \cQ_t=\mathcal{Q}, \ t\in\cT$, is dynamically consistent.
\end{proposition}
\proof

If $\mathcal{Q}\subseteq\mathcal{P}$ is strongly consistent, then, for every
$A\in\mathcal{F}_t, \ \mathcal{F}_t$-measurable random variable $X$, and $t\in\{0,\dots,T-1\}$, we have
\begin{align}
1_A\inf\limits_{Q\in\mathcal{Q}} \mathbb{E}_{Q}\big[ X |\mathcal{F}_{t} \big]&=1_A\inf\limits_{Q\in\mathcal{Q}} \mathbb{E}_Q\Big[
\inf\limits_{M\in\mathcal{Q}} \mathbb{E}_M \big[   X  \, |\, \mathcal{F}_{t+1} \big] \, | \, \mathcal{F}_t \Big] \nonumber \\
&=1_A\inf\limits_{Q\in\mathcal{Q}} \mathbb{E}_Q\Big[1_A
\inf\limits_{M\in\mathcal{Q}} \mathbb{E}_M \big[   X  \, |\, \mathcal{F}_{t+1} \big] \, | \, \mathcal{F}_t \Big] \nonumber \\
&\leq 1_A\inf\limits_{Q\in\mathcal{Q}} \mathbb{E}_Q\Big[1_A
\max_{\omega\in A}\bigg\{\inf\limits_{Q\in\mathcal{Q}} \mathbb{E}_{Q}\big[ X|\mathcal{F}_{t+1} \big](\omega)\bigg\} \, | \, \mathcal{F}_t \Big]\nonumber \\
&\leq 1_A\inf\limits_{Q\in\mathcal{Q}} \mathbb{E}_Q\Big[
\max_{\omega\in A}\bigg\{\inf\limits_{Q\in\mathcal{Q}} \mathbb{E}_{Q}\big[ X|\mathcal{F}_{t+1} \big](\omega)\bigg\} \, | \, \mathcal{F}_t \Big]\,. \label{eq:SetofProbs0}
\end{align}

Since $\max_{\omega\in A}\bigg\{\inf\limits_{Q\in\mathcal{Q}} \mathbb{E}_{Q}\big[ X|\mathcal{F}_{t+1} \big](\omega)\bigg\}$ is a constant, then, for each $Q\in\mathcal{Q}$, we have,
$$
\mathbb{E}_Q\Big[
\max_{\omega\in A}\bigg\{\inf\limits_{Q\in\mathcal{Q}} \mathbb{E}_{Q}\big[ X|\mathcal{F}_{t+1} \big](\omega)\bigg\} \, | \, \mathcal{F}_t \Big]=\max_{\omega\in A}\bigg\{\inf\limits_{Q\in\mathcal{Q}} \mathbb{E}_{Q}\big[ X|\mathcal{F}_{t+1} \big](\omega)\bigg\}\,.
$$
Therefore,
$$
\inf\limits_{Q\in\mathcal{Q}}\mathbb{E}_Q\Big[
\max_{\omega\in A}\bigg\{\inf\limits_{Q\in\mathcal{Q}} \mathbb{E}_{Q}\big[ X|\mathcal{F}_{t+1} \big](\omega)\bigg\} \, | \, \mathcal{F}_t \Big]=\max_{\omega\in A}\bigg\{\inf\limits_{Q\in\mathcal{Q}} \mathbb{E}_{Q}\big[ X|\mathcal{F}_{t+1} \big](\omega)\bigg\}\,.
$$
The last equality together with \eqref{eq:SetofProbs0} imply
$$
1_A\inf\limits_{Q\in\mathcal{Q}} \mathbb{E}_{Q}\big[ X |\mathcal{F}_{t} \big]\leq 1_A\max_{\omega\in A}\bigg\{\inf\limits_{Q\in\mathcal{Q}} \mathbb{E}_{Q}\big[ X|\mathcal{F}_{t+1} \big](\omega)\bigg\} \ .
$$
Finally note that for any set of probability measures $\mathcal{Q}\subseteq\mathcal{P}$, we have
\begin{align}
1_A\min_{\omega\in A}\bigg\{\inf_{\mathbb{Q}\in\mathcal{Q}}\mathbb{E}_{\mathbb{Q}}[X|\mathcal{F}_{t+1}](\omega)\bigg\}
& \leq 1_A\inf_{\mathbb{Q}\in\mathcal{Q}}\mathbb{E}_{\mathbb{Q}}[X|\mathcal{F}_{t}]\, , \label{eq:SetofProbs1-0}
\end{align}
for every $t\in\{0,\dots,T-1\}$, $A\in\mathcal{F}_t$,  and every random variable $X$.
\endproof

The rest of the subsection is dedicated to examples of dynamically consistent sequences of sets of probability measures.

\begin{example}\label{ex:Set-Prob-1}
Singleton set $\mathcal{Q}=\{Q\}$, with $\mathbb{Q}\in\cP^e$,  is clearly strongly consistent.
By Proposition \ref{prop:cons2DCons}  the constant sequence
$\{ \cQ, \cQ, \ldots,\cQ\}$ is dynamically consistent. For simplicity of writing,
we will denote this sequence by $\mathcal{Q}^{s}$.
\end{example}

\begin{example}\label{ex:Set-Prob-2}
It is not hard to show that
$$
\sum_{i=1}^{n_t}1_{P^t_i}\inf\limits_{\mathbb{Q}\in\mathcal{P}^e}\mathbb{E}_{\mathbb{Q}}
[D|\mathcal{F}_t]=\sum_{i=1}^{n_t}1_{P^t_i}\min\limits_{\omega\in P^t_i}D(\omega)\, , \quad t\in\cT, \ D\in\mathcal{D}.
$$
This implies that the set $\mathcal{P}^e$ of all equivalent probability measures with respective to $\mathbb{P}$, is consistent.
Hence, the constant sequence $\{\cP^e,\cP^e,\ldots,\cP^e\}$ is dynamically consistent.

\end{example}

\begin{example}\label{ex:Set-Prob-3}
Let $a\geq 1$ be a real number. The following set of probability measures
$$
\mathcal{Q}^{a,u}:=\{\mathbb{Q}\in\mathcal{P}^e \ | \ \mathbb{E}_{\mathbb{P}}[ d\mathbb{Q}/d\mathbb{P} |\mathcal{F}_t]
\leq a\mathbb{E}_{\mathbb{P}}[ d\mathbb{Q}/d\mathbb{P}  | \mathcal{F}_{t-1}] \textrm{ for all } t\in\{1,\ldots,T\} \}
$$
is consistent.
\end{example}
\noindent
First note that
$$
\mathop{\inf}\limits_{\widetilde{\mathbb{Q}}\in\mathcal{Q}^{a,u}}\mathbb{E}_{\widetilde{\mathbb{Q}}}[X|\mathcal{F}_t]
\geq \mathop{\inf}\limits_{\mathbb{Q}\in\mathcal{Q}^{a,u}}\mathbb{E}_{\mathbb{Q}}
[\mathop{\inf}\limits_{\mathbb{M}\in\mathcal{Q}^{a,u}}\mathbb{E}_{\mathbb{M}}[X|\mathcal{F}_{t+1}]|\mathcal{F}_t]\, ,
$$
for every $t\in\{0,\dots,T-1\}$ and $\mathcal{F}_t$-measurable bounded random variable $X$. Next we will show that
the converse inequality also holds true and hence, by definition, $\mathcal{Q}^{a,u}$ is
consistent.
Towards this end, assume that  $t\in\cT$, $X$ is an $\mathcal{F}_t$-measurable random variable, and $a\geq 1$; all arbitrary but fixed in what follows.
For convenience, we denote by $P^{t+1}_{i,j}$ the set of partition $(P^{t+1}_1,\dots,P^{t+1}_{n_{t+1}})$ such
that $P^t_i=\cup_{j=1}^{k_i} P^{t+1}_{i,j}, \ i=1,\ldots,n_t$.
Note that $k_1+k_2+\dots+k_{n_t}=n_{t+1}$.

Pick up arbitrarily $n_t+n_{t+1}$ probability measures from $\mathcal{Q}^{a,u}$, and denote them by
$(\mathbb{Q}_1,\mathbb{Q}_2$, $\dots,\mathbb{Q}_{n_t},\mathbb{M}_{1,1}$,
 $\mathbb{M}_{1,2},\dots,\mathbb{M}_{1,k_1},\mathbb{M}_{2,1}, \mathbb{M}_{2,2},\dots,\mathbb{M}_{2,k_2},\dots \dots,\mathbb{M}_{n_t,1},\mathbb{M}_{n_t,2}, \dots,\mathbb{M}_{n_t,k_{n_t}})$. Some of them are allowed to be the same.
We will construct a new probability measure based on the above set of probabilities.
For any $i\in \{1,2,\dots,n_t\}$, $j\in\{1,2,\dots,k_i\}$, and $\omega \in P^{t+1}_{i,j}$ we put
\begin{align*}
\mathbb{H}(\omega) :=
{\mathbb{M}_{i,j}(\omega)\over \mathbb{M}_{i,j}(P^{t+1}_{i,j})}{\mathbb{Q}_{i}(P^{t+1}_{i,j})\over \mathbb{Q}_{i}(P^{t}_{i})}\mathbb{P}(P^t_i)\,.
\end{align*}
Note that $P_{i,j}^{t+1}$, $i\in \{1,2,\dots,n_t\}$, $j\in\{1,2,\dots,k_i\}$, is a partition of $\Omega$, and hence $\mathbb{H}$ is well-defined,
and since all probability measures in $\cQ$ are of full support, $\mathbb{H}(\omega)$ is finite for all $\omega\in\Omega$.
It is also easy to show that $\mathbb{H}(\Omega)=1$, and thus $\mathbb{H}$ is a probability measure.

Next we will prove that $\mathbb{H}\in\mathcal{Q}^{a,u}$.
On any set $P^{t}_{i}$, we have
\begin{align*}
1_{P^t_i}\mathbb{E}_{\mathbb{P}}[{d\mathbb{H}\over d\mathbb{P}}|\mathcal{F}_t]
&=1_{P^t_i}\sum_{j=1}^{k_i}\sum_{\omega\in P^{t+1}_{i,j}}{\mathbb{H}(\omega)\over\mathbb{P}(P^t_i)}
=1_{P^t_i}\sum_{j=1}^{k_i}\sum_{\omega\in P^{t+1}_{i,j}}{\mathbb{M}_{i,j}(\omega)\over \mathbb{M}_{i,j}(P^{t+1}_{i,j})}{\mathbb{Q}_{i}(P^{t+1}_{i,j})\over \mathbb{Q}_{i}(P^{t}_{i})}{\mathbb{P}(P^t_i)\over\mathbb{P}(P^t_i)}\\
&=1_{P^t_i}\sum_{j=1}^{k_i}{\mathbb{M}_{i,j}(P^{t+1}_{i,j})\over \mathbb{M}_{i,j}(P^{t+1}_{i,j})}{\mathbb{Q}_{i}(P^{t+1}_{i,j})\over \mathbb{Q}_{i}(P^{t}_{i})}{\mathbb{P}(P^t_i)\over\mathbb{P}(P^t_i)}
=1_{P^t_i}\sum_{j=1}^{k_i}{\mathbb{Q}_{i}(P^{t+1}_{i,j})\over \mathbb{Q}_{i}(P^{t}_{i})}= 1_{P^t_i}\,.
\end{align*}
Thus,
$\mathbb{E}_{\mathbb{P}}[{d\mathbb{H}\over d\mathbb{P}}|\mathcal{F}_t]=1$, and by tower property, for all $s\leq t$, we also have
$\mathbb{E}_{\mathbb{P}}[{d\mathbb{H}\over d\mathbb{P}}|\mathcal{F}_s]=1$. Consequently, we get
\begin{equation}\label{eq:eq:SetofProbs3}
\mathbb{E}_{\mathbb{P}}[{d\mathbb{H}\over d\mathbb{P}}|\mathcal{F}_{s}]
 \leq a\mathbb{E}_{\mathbb{P}}[{d\mathbb{H}\over d\mathbb{P}}|\mathcal{F}_{s-1}], \quad  \textrm{for all } \ s\leq t\,.
\end{equation}
On the other hand, for any $P^{t+1}_{i,j}$, we have,
\begin{align*}
1_{P^{t+1}_{i,j}}\mathbb{E}_{\mathbb{P}}[{d\mathbb{H}\over d\mathbb{P}}|\mathcal{F}_{t+1}]
=1_{P^{t+1}_{i,j}}\sum_{\omega\in P^{t+1}_{i,j}}{\mathbb{H}(\omega)\over \mathbb{P}(P^{t+1}_{i,j})}
 =1_{P^{t+1}_{i,j}}{\mathbb{Q}_{i}(P^{t+1}_{i,j})\over \mathbb{Q}_{i}(P^{t}_{i})}{\mathbb{P}(P^t_i)\over\mathbb{P}(P^{t+1}_{i,j})}\,.
\end{align*}
Since $\mathbb{Q}_i\in\mathcal{Q}^{a,u}$, we have that
$\mathbb{E}_{\mathbb{P}}[{d\mathbb{Q}_i\over d\mathbb{P}}|\mathcal{F}_{t+1}]\leq a\mathbb{E}_{\mathbb{P}}[{d\mathbb{Q}_i\over d\mathbb{P}}|\mathcal{F}_{t}]$,
and thus $1_{P^{t+1}_{i,j}}\mathbb{E}_{\mathbb{P}}[{d\mathbb{Q}_i\over d\mathbb{P}}|\mathcal{F}_{t+1}]\leq a1_{P^{t+1}_{i,j}}\mathbb{E}_{\mathbb{P}}[{d\mathbb{Q}_i\over d\mathbb{P}}|\mathcal{F}_t]$.
This implies that
$1_{P^{t+1}_{i,j}}{\mathbb{Q}_i(P^{t+1}_{i,j})\over \mathbb{P}(P^{t+1}_{i,j})}\leq a1_{P^{t+1}_{i,j}}{\mathbb{Q}_i(P^{t}_{i})\over \mathbb{P}(P^{t}_{i})}$.
Hence,
${\mathbb{Q}_{i}(P^{t+1}_{i,j})\over \mathbb{P}(P^{t+1}_{i,j})} {\mathbb{P}(P^t_i)\over \mathbb{Q}_{i}(P^{t}_{i})}\leq a$,
and therefore,
\begin{align*}
1_{P^{t+1}_{i,j}}\mathbb{E}_{\mathbb{P}}[{d\mathbb{H}\over d\mathbb{P}}|\mathcal{F}_{t+1}]\leq 1_{P^{t+1}_{i,j}}a=1_{P^{t+1}_{i,j}}a\mathbb{E}_{\mathbb{P}}[{d\mathbb{H}\over d\mathbb{P}}|\mathcal{F}_t]\,.
\end{align*}
Since the above holds true for any $P^{t+1}_{i,j}$, we have that
$$
\mathbb{E}_{\mathbb{P}}[{d\mathbb{H}\over d\mathbb{P}}|\mathcal{F}_{t+1}]\leq a\mathbb{E}_{\mathbb{P}}[{d\mathbb{H}\over d\mathbb{P}}|\mathcal{F}_t]\,.
$$
By similar arguments as above, inductively, one can show that
$$
\mathbb{E}_{\mathbb{P}}[{d\mathbb{H}\over d\mathbb{P}}|\mathcal{F}_{s}]\leq a\mathbb{E}_{\mathbb{P}}[{d\mathbb{H}\over d\mathbb{P}}|\mathcal{F}_t]\, ,
$$
for any $s>t$. Combining this with \eqref{eq:eq:SetofProbs3}, we conclude that $\mathbb{H}\in\cQ^{a,u}$.

Next let us evaluate $\mathbb{E}_{\mathbb{H}}[D|\mathcal{F}_t]$.
Consider a new random variable $Y$, defined as follows:
$$
Y:=\sum_{i=1}^{n_t}\sum_{j=1}^{k_i}1_{P^{t+1}_{i,j}}\mathbb{E}_{\mathbb{M}_{i,j}}[D|\mathcal{F}_{t+1}]\,.
$$
Then, for any $m\in\{1,2,\dots,n_t\}$, we can deduce that
\begin{align}
1_{P^t_m}\mathbb{E}_{\mathbb{Q}_m}[Y|\mathcal{F}_t]
=1_{P^t_m}\sum_{j=1}^{k_m}\mathbb{E}_{\mathbb{Q}_m}[1_{P^{t+1}_{m,j}}\mathbb{E}_{\mathbb{M}_{m,j}} [D|\mathcal{F}_{t+1}]|\mathcal{F}_t]\,. \label{eq:SetofProbs4}
\end{align}
For convenience, we put
$C^{t+1}_{m,j} :=1_{P^{t+1}_{m,j}}\sum_{\omega\in P^{t+1}_{m,j}}{\mathbb{M}_{m,j}(\omega)D(\omega)\over \mathbb{M}_{m,j}(P^{t+1}_{m,j})}$.
Note that
$$
1_{P^{t+1}_{m,j}}\mathbb{E}_{\mathbb{M}_{m,j}} [D|\mathcal{F}_{t+1}]= C^{t+1}_{m,j}.
$$ Hence,
\begin{align*}
1_{P^t_m}\mathbb{E}_{\mathbb{Q}_m}[Y|\mathcal{F}_t]=
1_{P^t_m}\sum_{j=1}^{k_m}\mathbb{E}_{\mathbb{Q}_m}[C^{t+1}_{m,j}|\mathcal{F}_t] 
=\sum_{\bar{\omega}\in P^t_m}{\mathbb{Q}_m(\bar{\omega})\over\mathbb{Q}_m(P^t_m)}\left(\sum_{j=1}^{k_m}C^{t+1}_{m,j}(\bar{\omega})\right)\,.
\end{align*}
By the definition of $C^{t+1}_{m,j}$, we note that
$$
\sum_{j=1}^{k_m}C^{t+1}_{m,j}(\bar{\omega})=\sum_{j=1}^{k_m}\Big[1_{P^{t+1}_{m,j}}(\bar{\omega})\sum_{\omega\in P^{t+1}_{m,j}}{\mathbb{M}_{m,j}(\omega)\over \mathbb{M}_{m,j}(P^{t+1}_{m,j})}D(\omega)\Big]\,.
$$
Then,
\begin{align*}
1_{P^t_m}\mathbb{E}_{\mathbb{Q}_m}[Y|\mathcal{F}_t]&=\sum_{\bar{\omega}\in P^t_m}\Big[{\mathbb{Q}_m(\bar{\omega})\over\mathbb{Q}_m(P^t_m)}\sum_{j=1}^{k_m}
[1_{P^{t+1}_{m,j}}(\bar{\omega})\sum_{\omega\in P^{t+1}_{m,j}}{\mathbb{M}_{m,j}(\omega)\over \mathbb{M}_{m,j}(P^{t+1}_{m,j})}D(\omega)]\Big]\\
&=\sum_{u=1}^{k_m}\sum_{\omega\in P^{t+1}_{m,u}}{\mathbb{Q}_m(P^{t+1}_{m,u})\over\mathbb{Q}_m(P^t_m)}
{\mathbb{M}_{m,u}
(\omega)\over \mathbb{M}_{m,u}(P^{t+1}_{m,u})}D(\omega) .
\end{align*}
From here, using the fact that $\mathbb{H}(P^t_i)=\mathbb{P}(P^t_i)$, we conclude that
\begin{align*}
1_{P^t_m}\mathbb{E}_{\mathbb{Q}_m}[Y|\mathcal{F}_t]
=1_{P^t_m}\mathbb{E}_{\mathbb{H}}[D|\mathcal{F}_t]\,.
\end{align*}
Since $\mathbb{H}\in\mathcal{Q}^{a,u}$, we have that $\mathbb{E}_{\mathbb{H}}[D|\mathcal{F}_t]
\geq \inf\limits_{\widetilde{\mathbb{Q}}\in\mathcal{Q}^{a,u}}\mathbb{E}_{\widetilde{\mathbb{Q}}}[D|\mathcal{F}_t]$. Consequently,
the following inequality holds true
$$
1_{P^t_m}\mathbb{E}_{\mathbb{Q}_m}[Y|\mathcal{F}_t]\geq 1_{P^t_m}\inf\limits_{\widetilde{\mathbb{Q}}\in\mathcal{Q}^{a,u}}
\mathbb{E}_{\widetilde{\mathbb{Q}}}[D|\mathcal{F}_t]\,.
$$
By \eqref{eq:SetofProbs4}, it follows that
$$
1_{P^t_m}\mathbb{E}_{\mathbb{Q}_m}[Y|\mathcal{F}_t]=1_{P^t_m}\mathbb{E}_{\mathbb{Q}_m}[\sum_{i=1}^{n_t}\sum_{j=1}^{k_i}
1_{P^{t+1}_{i,j}}\mathbb{E}_{\mathbb{M}_{i,j}}[D|\mathcal{F}_{t+1}]|\mathcal{F}_t],
$$
from which we continue
$$
1_{P^t_m}\mathbb{E}_{\mathbb{Q}_m}[Y|\mathcal{F}_t] \geq
1_{P^t_m}\mathbb{E}_{\mathbb{Q}_m}\Big[\sum_{i=1}^{n_t}\sum_{j=1}^{k_i}
1_{P^{t+1}_{i,j}}\inf_{\mathbb{M}_{i,j}\in\mathcal{Q}^{a,u}}\mathbb{E}_{\mathbb{M}_{i,j}}
[D|\mathcal{F}_{t+1}]|\mathcal{F}_t\Big]
\geq 1_{P^t_m}\inf\limits_{\widetilde{\mathbb{Q}}\in\mathcal{Q}^{a,u}}
\mathbb{E}_{\widetilde{\mathbb{Q}}}[D|\mathcal{F}_t]\, ,
$$
and since, this is true for all $\mathbb{Q}_m\in\mathcal{Q}^{x,u}$, we have
$$
1_{P^t_m}\inf_{\mathbb{Q}_m\in\mathcal{Q}^{a,u}}\mathbb{E}_{\mathbb{Q}_m}\Big[\sum_{i=1}^{n_t}\sum_{j=1}^{k_i}
1_{P^{t+1}_{i,j}}\inf_{\mathbb{M}_{i,j}\in\mathcal{Q}^{a,u}}\mathbb{E}_{\mathbb{M}_{i,j}}
[D|\mathcal{F}_{t+1}]|\mathcal{F}_t\Big]
\geq 1_{P^t_m}\inf\limits_{\widetilde{\mathbb{Q}}\in\mathcal{Q}^{a,u}}
\mathbb{E}_{\widetilde{\mathbb{Q}}}[D|\mathcal{F}_t] \, .
$$
Summing both sides of the last inequality over $m\in\{1,2,\dots,n_t\}$, we have
$$
\sum_{m=1}^{n_t}1_{P^t_m}\inf_{\mathbb{Q}_m\in\mathcal{Q}^{a,u}}\mathbb{E}_{\mathbb{Q}_m}\Big[\sum_{i=1}^{n_t}\sum_{j=1}^{k_i}
1_{P^{t+1}_{i,j}}\inf_{\mathbb{M}_{i,j}\in\mathcal{Q}^{a,u}}\mathbb{E}_{\mathbb{M}_{i,j}}
[D|\mathcal{F}_{t+1}]|\mathcal{F}_t\Big]
\geq \sum_{m=1}^{n_t}1_{P^t_m}\inf\limits_{\widetilde{\mathbb{Q}}\in\mathcal{Q}^{a,u}}
\mathbb{E}_{\widetilde{\mathbb{Q}}}[D|\mathcal{F}_t] ,
$$
or equivalently,
\begin{align*}
\mathop{\inf}\limits_{\mathbb{Q}\in\mathcal{Q}^{a,u}}\mathbb{E}_{\mathbb{Q}}
\Big[\mathop{\inf}\limits_{\mathbb{M}\in\mathcal{Q}^{a,u}}\mathbb{E}_{\mathbb{M}}[D|\mathcal{F}_{t+1}]|\mathcal{F}_t\Big]
\geq\mathop{\inf}\limits_{\widetilde{\mathbb{Q}}\in\mathcal{Q}^{a,u}}\mathbb{E}_{\widetilde{\mathbb{Q}}}[D|\mathcal{F}_t]\, .
\end{align*}
This concludes the proof that $\cQ^{a,u}$ is consistent.

\begin{remark}
It is easy to show that for any $\mathbb{Q}\in\mathcal{Q}^{a,u}$,
$$
\mathbb{E}_{\mathbb{P}}[{d\mathbb{Q}\over d\mathbb{P}}|\mathcal{F}_t]\leq a^t\, , \quad t\in\cT.
$$
In particular, $\mathbb{Q}(A)\leq a^t\mathbb{P}(A)$, for any $\mathbb{Q}\in\cQ^{a,u}, \ A\in\mathcal{F}_t$ and $t\in\cT$. 
Different probabilities in $\cQ^{a,u}$ can be regarded as different opinions about the distribution of cash-flows; the above inequality provides an upper bound of these probabilities in terms of the underlying probability $\cP$.
\end{remark}

\begin{example}\label{ex:Set-Prob-4} By similar arguments as in previous examples, one can show that the set of probability measures
$\cQ^{a,l}$ defined as follows
$$
\mathcal{Q}^{a,l} :=\{\mathbb{Q}\in\mathcal{P}^e \ | \ \mathbb{E}_{\mathbb{Q}}[d\mathbb{P}/d\mathbb{Q} \, |\, \mathcal{F}_j]
\leq a\mathbb{E}_{\mathbb{Q}}[ d\mathbb{P}/d\mathbb{Q} \, |\, \mathcal{F}_{j-1}] \textrm{ for all } j=1,\dots,T, \}
$$
is a  consistent set of probability measures.
\end{example}

\begin{example}\label{ex:Set-Prob-5} In this example we construct a dynamically consistent sequence of sets probability  measures that is not constant sequence of consistent sets of probability measures.

Let $\bP_0, \bP_1,\ldots,\bP_T$, be a sequence of probability measures in $\cP^e$ such that $\bP_i\neq\bP_j$, for $i\neq j$. Consider the following sequence of sets of probability measures $\cQ_t=\cP\setminus\bP_t, \ t=0,1,\ldots,T$. It is easy to show that
\begin{equation}\label{eq:Ex-Prob-Sets5}
\inf_{\mathbb{Q}\in\mathcal{Q}_{t}}\mathbb{E}_{\mathbb{Q}}[X|\mathcal{F}_{t}]
=\inf_{\mathbb{Q}\in\mathcal{P}^e}\mathbb{E}_{\mathbb{Q}}[X|\mathcal{F}_{t}], \quad t\in\cT.
\end{equation}
This implies that $\{\cQ_t\}_{t=0}^T$ is a dynamically consistent sequence of sets of probabilities measures.  Clearly it is not a constant sequence.

\end{example}

\subsection{Representation Theorem of DCRM}\label{subsec:reprThforDCRM}
In this section we will present a representation theorem for dynamic coherent risk measures in terms of
dynamically consistent set of probabilities. These results combined with the results from Section \ref{subsec:dualityDCAIvsDCRM}
about duality between DCAI and DCRM will lead to a representation theorem
for dynamic coherent acceptability indices.

\begin{theorem}[Representation Theorem for DCRM]\label{th:RM-Representation}
A function $\rho: \{0,1,\ldots,T\}\times\mathcal{D}\times\Omega\to \mathbb{R}$ is a dynamic coherent risk measure
if and only if there exists a dynamically consistent family of sets of probabilities $\mathcal{U}:=\{\mathcal{Q}_s\}^T_{s=0}$ such that,
\begin{equation}\label{eq:RM-Representation}
\rho_t(D)=-\inf_{\mathbb{Q}\in\mathcal{Q}_t}\mathbb{E}_{\mathbb{Q}}\big[\sum_{s=t}^T D_s|\mathcal{F}_t\big]\, , \quad
\textrm{for all} \  t\in\cT, \ D\in\mathcal{D}.
\end{equation}

\end{theorem}
\proof {\bf Sufficiency}.
It is not hard to show that $\rho$ defined in \eqref{eq:RM-Representation} is a dynamic coherent risk measure.
(A1)-(A6) are checked similarly as in existing literature (see for instance \cite{Riedel2004}), and
for interest of saving space we will not check them here.
We will show only that (A7), dynamic consistency, is satisfied.

\noindent
Since $\mathcal{U}=\{\mathcal{Q}_t\}^T_{t=0}$ is dynamically consistent, we have,
\begin{align*}
1_A\rho_t(D)&=-1_A\inf_{\mathbb{Q}\in\mathcal{Q}_t}\mathbb{E}_{\mathbb{Q}}\big[\sum_{s=t}^T D_s|\mathcal{F}_t\big]
\geq 1_A\min_{\omega\in A}\bigg\{-\inf_{\mathbb{Q}\in\mathcal{Q}_{t+1}}\mathbb{E}_{\mathbb{Q}}[\sum_{s=t+1}^T D_s|\mathcal{F}_{t+1}](\omega)-D_t\bigg\}\\
&= 1_A\min_{\omega\in A}\bigg\{\rho_{t+1}(D,\omega)-D_t\bigg\}\,,
\end{align*}
for any $A\in\mathcal{F}_t, \ t\in\cT, \ D\in\cD$, and  $\mathcal{Q}_t\in\cU$.

Similarly, one can show that
$ 1_A\rho_t(D)= 1_A\max_{\omega\in A}\left\{\rho_{t+1}(D,\omega)-D_t\right\}$,
for every $t\in\cT, \ D\in\cD, \ \mathcal{Q}_t\in\cU$.
Thus (A7) is satisfied.

\smallskip \noindent
{\bf Necessity.} The set $\cU$ will be constructed explicitly.
Fix a time $t\in\cT$. Recall that
$\{P^t_1,\dots,P^t_{n_t}\}$ denotes the partition of $\Omega$ that corresponds to  $\cF_t$.
Also, we will denote by $\{P^{t,s}_{i,1},\dots,P^{t,s}_{i,m_s}\}$ the partition of $P_i^t$
generated by $\cF_s$, for some future time $s\geq t$. Thus $P^t_i=\cup_{j=1}^{m_s}P^{t,s}_{i,j}$.
Assume that $P_i^t$ is fixed for some $i\in\{1,\ldots,n_t\}$, and define the following probability space
$(\Omega^t_i,2^{\Omega^t_i},\mathbb{P}^{\textrm{uni}})$ with,
$$
\Omega^t_i:=\big\{(s,P^{t,s}_{i,j}): s\in\{t,t+1,\dots,T\} \textrm{ and } j\in\{1,2,\dots,m_s\} \big\}\, ,
$$
and $\mathbb{P}^{\textrm{uni}}(\omega) = 1/\textrm{card}(\Omega^t_i)$ for each $\omega \in\Omega^t_i$.

Let us denote by $\mathcal{X}(\Omega^t_i)$ the set of all random variables on $\Omega^t_i$.
There exists a one-to-one correspondence between $\mathcal{X}(\Omega^t_i)$
and the set
$\mathcal{D}_i^t:=\{D1_{\{t,t+1,\dots,T\}}1_{P^t_i}: \textrm{ for all } D\in\mathcal{D}\}$.
The map can be defined as follows: for any $X\in\mathcal{X}(\Omega^t_i)$, put
\begin{equation}\label{proof:RM-Representation-2}
D^X_s(\omega):=\left\{ \begin{array}{cc}
         X((s,P^{t,s}_{i,j})), & \textrm{ if } s\geq t \textrm{ and } \omega\in P^{t,s}_{i,j} \\
                  0, & \textrm{ otherwise} \, ,
                          \end{array} \right.
\end{equation}
and vise versa, for any $D\in\mathcal{D}_i^t$, define
\begin{equation}\label{proof:RM-Representation-3}
X^D((s,P^{t,s}_{i,j})):=D_s(\omega),
\end{equation}
for $s\geq t$, $j\in\{1,2,\dots,m_s\}$, and $\omega\in P^{t,s}_{i,j}$.

Consider the following map $\phi:\mathcal{X}(\Omega^t_i)\to \mathbb{R}$ with,
\begin{align}\label{proof:RM-Representation-1}
\phi(X):={1\over T-t+1}\rho_t(D^X,\omega), \quad \omega\in P_i^t.
\end{align}
We claim that $\phi$ is a static coherent risk measure, i.e. satisfies the properties (R1)-(R4) of Definition \ref{d23}.
Indeed, for any $X,Y\in\mathcal{X}(\Omega^t_i)$, such that $X\leq Y$, we have,
$D^X_s(\omega)\leq D^Y_s(\omega)$,
for all $s\geq t$ and $\omega\in \Omega$. Then, by (A3), the monotonicity of $\rho$, we get
$\rho_t(D^X,\omega)\geq \rho_t(D^Y,\omega)$, for $\omega\in\Omega$. Therefore, by \eqref{proof:RM-Representation-1},
$\phi(X)\geq \phi(Y)$, i.e. $\phi$ satisfies (R1).

Note that for all $X\in\mathcal{X}(\Omega^t_i)$ and $\lambda\geq 0$, by \eqref{proof:RM-Representation-2}, we have,
\begin{align*}
D^{\lambda X}_s(\omega)=\lambda X((s,P^{t,s}_{i,j}))=\lambda D^{X}_s(\omega)\,,
\end{align*}
for all $s\geq t$ and $\omega\in P^{t,s}_{i,j}$. From here, by
\eqref{proof:RM-Representation-1} and using homogeneity of $\rho$, the homogeneity (R2) of $\phi$ follows.

Next we will show that $\phi$ satisfies (R3).
For all $X\in\mathcal{X}(\Omega^t_i)$ and $k\in\mathbb{R}$, by \eqref{proof:RM-Representation-2}, we have,
\begin{align*}
D^{X+k}_s(\omega)=X((s,P^{t,s}_{i,j}))+k=D^{X}_s(\omega)+k\,,
\end{align*}
for all $s\geq t$ and $\omega\in P^{t,s}_{i,j}$.
Therefore, by \eqref{proof:RM-Representation-1} and (A6), translation invariance of $\rho$,  we deduce
\begin{align*}
\phi(X+k)&= {1\over T-t+1}\rho_t(D^X+k1_{\{t,\dots,T\}},\omega) \\
&={1\over T-t+1}(\rho_t(D^X,\omega)-(T-t+1)k) \\
&=\phi(X^{D})-k\,,
\end{align*}
for all $X\in\mathcal{X}(\Omega^t_i)$.

To show that $\phi$ satisfies (R4), consider an $X\in\mathcal{X}(\Omega^t_i)$. By
\eqref{proof:RM-Representation-2}
$D^{X+Y}_s(\omega)=D^{X}_s(\omega)+D^{Y}_s(\omega)$,
 for all $s\geq t$ and $\omega\in P^{t,s}_{i,j}$,
and therefore, by \eqref{proof:RM-Representation-1} and (A5), subadditivity of $\rho$, we obtain
\begin{align*}
\phi(X+Y)&={1\over T-t+1}\rho_t(D^{X}+D^{Y},\omega)
\leq{1\over T-t+1}\left( \rho_t(D^X,\omega)+\rho_t(D^Y,\omega)\right)\\
&=\phi(X)+\phi(Y)\, .
\end{align*}

From all the above, we conclude that $\phi$ is a static coherent risk measure.
By Theorem \ref{th:ReprThStaticCRM}, representation of static coherent risk measures,
there exists $\mathcal{M}^t_i$, a set of absolutely continuous probability measures with respect to $\mathbb{P}^{\textrm{uni}}$
on $\Omega^t_i$, such that
$$
\phi(X)=-\inf_{\mathbb{M}\in\mathcal{M}^t_i}\mathbb{E}_{\mathbb{M}}[X]\,.
$$
By \eqref{proof:RM-Representation-1}, we have,
\begin{align}\label{proof:RM-Representation-4}
{1\over T-t+1}\rho_t(D^X,\omega)=-\inf_{\mathbb{M}\in\mathcal{M}^t_i}\mathbb{E}_{\mathbb{M}}[X]\,,  \quad \omega\in P^t_i.
\end{align}
Since there is one-to-one map between $\mathcal{X}(\Omega^t_i)$ and $\mathcal{D}^t_i$, for any $D\in\mathcal{D}^t_i$, we also can write
\begin{align}\label{proof:RM-Representation-5}
{1\over T-t+1}\rho_t(D,\omega)=-\inf_{\mathbb{M}\in\mathcal{M}^t_i}\mathbb{E}_{\mathbb{M}}[X^D]\,.
\end{align}

Fix a time $t^0\in\{t,t+1,\dots,T\}$, and denote by $\widetilde{D}$ the process $1_{\{t^0\}}$.
By (A6)-translation invariance and (A2)-independence of the past of $\rho$,
it follows that $\rho_t(\widetilde{D},\omega)=-1$, $\omega\in P^t_i$.
Hence, by \eqref{proof:RM-Representation-5},
\begin{align}\label{proof:RM-Representation-6}
\inf_{\mathbb{M}\in\mathcal{M}^t_i}\mathbb{E}_{\mathbb{M}}[X^{\widetilde{D}}]
= \frac{1}{T-t+1}\, .
\end{align}
Note that $\mathbb{E}_{\mathbb{M}}[X^{\widetilde{D}}]=\mathbb{M}(\{t^0\}\times P^t_i)$.
Thus, \eqref{proof:RM-Representation-6} implies
$$
\inf_{\mathbb{M}\in\mathcal{M}^t_i}\mathbb{M}(\{t^0\}\times P^t_i)={1\over T-t+1}\,.
$$
Similarly, one can show that
$\mathbb{E}_{\mathbb{M}}[X^{-\widetilde{D}}] =-\mathbb{M}(\{t^0\}\times P^t_i)$.
Thus we derive that
$$
\inf_{\mathbb{M}\in\mathcal{M}^t_i}\mathbb{E}_{\mathbb{M}}[X^{-\widetilde{D}}]
=\inf_{\mathbb{M}\in\mathcal{M}^t_i}(-\mathbb{M}(\{t^0\}\times P^t_i))=-\sup_{\mathbb{M}\in\mathcal{M}^t_i}\mathbb{M}(\{t^0\}\times P^t_i)\, ,
$$
and consequently
$$
\sup_{\mathbb{M}\in\mathcal{M}^t_i}\mathbb{M}(\{t^0\}\times P^t_i)={1\over T-t+1}\,.
$$
This yields that
\begin{align}\label{proof:RM-Representation-8}
\mathbb{M}(\{t^0\}\times P^t_i)={1\over T-t+1}\,,\quad t^0\in\{t,t+1,\dots,T\}.
\end{align}
For any $s\in\{t,t+1,\dots,T\}$, define $\mathbb{M}^s\,:\,\Omega^t_i\to \mathbb{R}$ as follows
$$
\mathbb{M}^s((r,P^{t,r}_{i,j})):=
\begin{cases}
(T-t+1)\mathbb{M}((r,P^{t,r}_{i,j})), & \textrm{ when }r=s \textrm{ and } j \in\{1,2,\dots,m_r\} \\
              0, & \textrm{ otherwise} .
\end{cases}
$$
It is straightforward to show that $\mathbb{M}^s$ is a probability measure on $\Omega^t_i$ for every $s\in\{t,t+1,\dots,T\}$.

For all $D\in\mathcal{D}$, we can derive,
\begin{align*}
\sum_{s=t}^T \mathbb{E}_{\mathbb{M}^s}[X^{D_s1_s}]
&=\sum_{s=t}^T\big(\sum_{r=t}^T\sum_{j=1}^{m_r}\mathbb{M}^s((r,P^{t,r}_{i,j}))(D_s1_{s})_r(\omega)\big), \textrm{ for some } \omega\in P^{t,r}_{i,j} \\
&=\sum_{s=t}^T\big(\sum_{j=1}^{m_r}\mathbb{M}^s((s,P^{t,s}_{i,j}))D_s(\omega)\big), \textrm{ for some } \omega\in P^{t,r}_{i,j} \\
&=\sum_{s=t}^T\big(\sum_{j=1}^{m_r}(T-t+1)\mathbb{M}((s,P^{t,s}_{i,j}))D_s(\omega)\big), \textrm{ for some } \omega\in P^{t,r}_{i,j} \\
&=(T-t+1)\mathbb{E}_{\mathbb{M}}[X^D]\,.
\end{align*}
Hence, by \eqref{proof:RM-Representation-5}, we have
\begin{equation}
\rho_t(D,\omega)=-(T-t+1)\inf_{\mathbb{M}\in\mathcal{M}^t_i}\mathbb{E}_{\mathbb{M}}[X^D]
=-\inf_{\mathbb{M}\in\mathcal{M}^t_i}\sum_{s=t}^T \mathbb{E}_{\mathbb{M}^s}[X^{D_s1_{s}}],  \quad \omega\in P^t_i. \label{proof:RM-Representation-9}
\end{equation}
Since $\rho$ satisfies (A6) and (A7), we deduce that
$$
\rho_s(D_s1_{\{s\}}-D_s1_{\{T\}},\omega)=0\,, \quad s\geq t, \ D\in\cD, \ \omega\in P_i^t.
$$
Thus, \eqref{proof:RM-Representation-5} and \eqref{proof:RM-Representation-9} imply,
\begin{align*}
-\inf_{\mathbb{M}\in\mathcal{M}^t_i}\left(\mathbb{E}_{\mathbb{M}^s}[X^{D_s1_{\{s\}}}]
-\mathbb{E}_{\mathbb{M}^T}[X^{D_s1_{\{T\}}}]\right)
=& -\inf_{\mathbb{M}\in\mathcal{M}^t_i}\left[\sum_{r=t}^T \mathbb{E}_{\mathbb{M}^r}[X^{(D_s1_{\{s\}}-D_s1_{\{T\}})_r1_r}]\right]\\
=& \rho_t(D_s1_{\{s\}}-D_s1_{\{T\}},\omega) = 0 \,.
\end{align*}
Since the above equality holds true for all $D\in\mathcal{D}$, it also holds true for $-D$.
Hence, we have
\begin{equation}\label{proof:RM-Representation-9-1}
\inf_{\mathbb{M}\in\mathcal{M}^t_i}(\mathbb{E}_{\mathbb{M}^s}[X^{-D_s1_{\{s\}}}]-\mathbb{E}_{\mathbb{M}^T}[X^{-D_s1_{\{T\}}}])=0 \,.
\end{equation}
On the other hand, by \eqref{proof:RM-Representation-3}, one gets
$$
\inf_{\mathbb{M}\in\mathcal{M}^t_i}(\mathbb{E}_{\mathbb{M}^s}[X^{-D_s1_{\{s\}}}]-\mathbb{E}_{\mathbb{M}^T}[X^{-D_s1_{\{T\}}}])
=-\sup_{\mathbb{M}\in\mathcal{M}^t_i}(\mathbb{E}_{\mathbb{M}^s}[X^{D_s1_{\{s\}}}]-\mathbb{E}_{\mathbb{M}^T}[X^{D_s1_{\{T\}}}])\,.
$$
Thus,
\begin{equation}\label{proof:RM-Representation-9-2}
\sup_{\mathbb{M}\in\mathcal{M}^t_i}(\mathbb{E}_{\mathbb{M}^s}[X^{D_s1_{\{s\}}}]-\mathbb{E}_{\mathbb{M}^T}[X^{D_s1_{\{T\}}}])=0
\end{equation}
By \eqref{proof:RM-Representation-9-1} and \eqref{proof:RM-Representation-9-2} we conclude that
$$
\sup_{\mathbb{M}\in\mathcal{M}^t_i}(\mathbb{E}_{\mathbb{M}^s}[X^{D_s1_{\{s\}}}]-\mathbb{E}_{\mathbb{M}^T}[X^{D_s1_{\{T\}}}])=0=
\inf_{\mathbb{M}\in\mathcal{M}^t_i}(\mathbb{E}_{\mathbb{M}^s}[X^{D_s1_{\{s\}}}]-\mathbb{E}_{\mathbb{M}^T}[X^{D_s1_{\{T\}}}]) \, ,
$$
and hence
\begin{align}\label{proof:RM-Representation-10}
\mathbb{E}_{\mathbb{M}^s}[X^{D_s1_{\{s\}}}]=\mathbb{E}_{\mathbb{M}^T}[X^{D_s1_{\{T\}}}]\,.
\end{align}
for all $s\geq t$, and $\mathbb{M}\in\mathcal{M}^t_i$.
Therefore, we can rewrite \eqref{proof:RM-Representation-9} as follows,
\begin{align}
\rho_t(D,\omega)&=-\inf_{\mathbb{M}\in\mathcal{M}^t_i}\Big[\sum_{s=t}^T \mathbb{E}_{\mathbb{M}^s}[X^{D_s1_{\{s\}}}]\Big]\nonumber\\
&=-\inf_{\mathbb{M}\in\mathcal{M}^t_i}\Big[\mathbb{E}_{\mathbb{M}^T}[\sum_{s=t}^T X^{D_s1_{\{T\}}}]\Big]\nonumber\\
&=-\inf_{\mathbb{M}\in\mathcal{M}^t_i}\mathbb{E}_{\mathbb{M}^T}\Big[X^{(\sum_{s=t}^T D_s)1_{\{T\}}}\Big]\label{proof:RM-Representation-11}\,.
\end{align}
for all $D\in\cD$, and $\omega\in P^t_i$.

To summarize, for every $P_i^t, \ i=1,\ldots,n_t$,
we constructed a set of probability measures $\mathcal{M}_i^t$ on $\Omega^t_i$.
Having these sets, we define $\mathcal{Q}_t$ as follows:
\begin{align*}
\mathcal{Q}_t\,:=\,\bigg\{\mathbb{Q} \in\cP \,:\,&\textrm{ there exists }\{\mathbb{M}_i\}^{n_t}_{i=1} \textrm{ such that, for all } i\in\{1,\dots,n_t\},
 \ j\in\{1,\dots,m^i_T\},\\
&\mathbb{M}_i\in\mathcal{M}_i^t \textrm{ and } \mathbb{Q}(\omega)={1\over n_t}{1\over \mathcal{N}(P^{t,T}_{i,j})}\mathbb{M}^T_i((T,P^{t,T}_{i,j})) \textrm{ for all } \omega\in P^{t,T}_{i,j} \bigg\}\, ,
\end{align*}
where $\mathcal{N}(P)$ stands for cardinality of the set $P\subset\Omega$.

By direct evaluations, one can show that $\cQ_t, \ t\in\cT$, is a set of probability measure on $\Omega$.

Next we will show that \eqref{eq:RM-Representation} is fulfilled.
Note that, for all $\omega \in P^t_i$,
\begin{align*}
\mathbb{E}_{\mathbb{Q}}\big[\sum_{s=t}^T D_s|\mathcal{F}_t\big](\omega)&=\sum_{\omega\in P^t_i}\bigg[\sum_{s=t}^T D_s(\omega)
{\mathbb{Q}(\omega)\over \mathbb{Q}(P^t_i)}\bigg]\\
&=\sum_{j=1}^{m_T^i}\sum_{\omega\in P^{t,T}_{i,j}}\bigg[\sum_{s=t}^T D_s(\omega)
{1\over \mathcal{N}(P^{t,T}_{i,j})}\mathbb{M}^T_i((T,P^{t,T}_{i,j}))\bigg]\\
&=\sum_{j=1}^{m_T^i}\big[\sum_{s=t}^T D_s(\omega)\mathbb{M}^T_i((T,P^{t,T}_{i,j}))\big] \\
&=\mathbb{E}_{\mathbb{M}^T_i}[X^{\sum_{s=t}^T D_s1_{\{T\}}}]\,.
\end{align*}
If $\inf_{\mathbb{Q}\in\mathcal{Q}_t}\mathbb{E}_{\mathbb{Q}}\big[\sum_{s=t}^T D_s|\mathcal{F}_t\big](\omega)>\inf_{\mathbb{M}_i\in\mathcal{M}^t_i}\mathbb{E}_{\mathbb{M}_i^T}[X^{\sum_{s=t}^T D_s1_{\{T\}}}]$,
then there exists $\widetilde{\mathbb{M}_i}\in\mathcal{M}_i^t$ such that
\begin{equation}\label{proof:RM-Representation-11-1}
\mathbb{E}_{\widetilde{\mathbb{M}}_i^T}[X^{\sum_{s=t}^T D_s1_{\{T\}}}]<\inf_{\mathbb{Q}\in\mathcal{Q}_t}\mathbb{E}_{\mathbb{Q}}\big[\sum_{s=t}^T D_s|\mathcal{F}_t\big](\omega)\,.
\end{equation}
However, for $\widetilde{\mathbb{Q}}$ constructed by $\widetilde{\mathbb{M}_i}$, as previously proved,
\begin{align*}
\mathbb{E}_{\widetilde{\mathbb{M}}^T_i}[X^{\sum_{s=t}^T D_s1_{\{T\}}}]&=\mathbb{E}_{\widetilde{\mathbb{Q}}}\big[\sum_{s=t}^T D_s|\mathcal{F}_t\big](\omega)
\geq \inf_{\mathbb{Q}\in\mathcal{Q}_t}\mathbb{E}_{\mathbb{Q}}\big[\sum_{s=t}^T D_s|\mathcal{F}_t\big](\omega)\,, \ \omega\in P_i^t,
\end{align*}
that contradicts \eqref{proof:RM-Representation-11-1}.
By the same arguments, one can show that the inequality
$$
\inf_{\mathbb{Q}\in\mathcal{Q}_t}\mathbb{E}_{\mathbb{Q}}\big[\sum_{s=t}^T D_s|\mathcal{F}_t\big](\omega)<\inf_{\mathbb{M}_i\in\mathcal{M}^t_i}\mathbb{E}_{\mathbb{M}_i^T}[X^{\sum_{s=t}^T D_s1_{\{T\}}}]\, ,
$$
can not hold true, and thus, we conclude that
\begin{align*}
\inf_{\mathbb{Q}\in\mathcal{Q}_t}\mathbb{E}_{\mathbb{Q}}\big[\sum_{s=t}^T D_s|\mathcal{F}_t\big](\omega)&=\inf_{\mathbb{M}_i\in\mathcal{M}^t_i}\mathbb{E}_{\mathbb{M}^T_i}[X^{\sum_{s=t}^T D_s1_{\{T\}}}]\,, \quad \omega\in P_i^t,
\end{align*}
and by \eqref{proof:RM-Representation-11},
$$
\rho_t(D)=-\inf_{\mathbb{Q}\in\mathcal{Q}_t}\mathbb{E}_{\mathbb{Q}}\big[\sum_{s=t}^T D_s|\mathcal{F}_t\big]\,.
$$

To complete the proof we need to show that $\{\mathcal{Q}_s\}^T_{s=0}$ is a dynamically consistent sequence of sets of probability measures.
Recall that by (A7), dynamic consistency of $\rho$,
\begin{equation}\label{proof:RM-Representation-12-1}
1_A (\min_{\omega\in A}\rho_{t+1}(D,\omega)-D_t) \leq 1_A \rho_t(D) \leq 1_A (\max_{\omega\in A}\rho_{t+1}(D,\omega)-D_t)\, ,
\end{equation}
for all $D\in\mathcal{D}$ and $A\in\mathcal{F}_t$. Using this, we get
$$
1_A (\min_{\omega\in A}\big\{-\inf_{\mathbb{Q}\in\mathcal{Q}_{t+1}}\mathbb{E}_{\mathbb{Q}}\big[\sum_{s=t+1}^T D_s|\mathcal{F}_{t+1}\big](\omega)\big\}-D_t )
\leq 1_A (-\inf_{\mathbb{Q}\in\mathcal{Q}_t}\mathbb{E}_{\mathbb{Q}}\big[\sum_{s=t}^T D_s|\mathcal{F}_t\big])\, ,
$$
for any $D\in\mathcal{D}$ and $A\in\cF_t$. Consequently, we obtain
\begin{equation}\label{proof:RM-Representation-12-2}
1_A \max_{\omega\in A}\big\{\inf_{\mathbb{Q}\in\mathcal{Q}_{t+1}}\mathbb{E}_{\mathbb{Q}}\big[\sum_{s=t}^T D_s|\mathcal{F}_{t+1}\big](\omega)\big\}
\geq 1_A\inf_{\mathbb{Q}\in\mathcal{Q}_t}\mathbb{E}_{\mathbb{Q}}\big[\sum_{s=t}^T D_s|\mathcal{F}_t\big]\,,
\ D\in\cD, \ A\in\cF_t.
\end{equation}
Similarly, by \eqref{proof:RM-Representation-12-1}
$$
1_A (\max_{\omega\in A}\big\{-\inf_{\mathbb{Q}\in\mathcal{Q}_{t+1}}\mathbb{E}_{\mathbb{Q}}\big[\sum_{s=t+1}^T D_s|\mathcal{F}_{t+1}\big](\omega)\big\}-D_t )
\geq 1_A (-\inf_{\mathbb{Q}\in\mathcal{Q}_t}\mathbb{E}_{\mathbb{Q}}\big[\sum_{s=t}^T D_s|\mathcal{F}_t\big])
$$
and hence
\begin{equation}\label{proof:RM-Representation-12-3}
1_A \min_{\omega\in A}\big\{\inf_{\mathbb{Q}\in\mathcal{Q}_{t+1}}\mathbb{E}_{\mathbb{Q}}\big[\sum_{s=t}^T D_s|\mathcal{F}_{t+1}\big](\omega)\big\}
\leq 1_A\inf_{\mathbb{Q}\in\mathcal{Q}_t}\mathbb{E}_{\mathbb{Q}}\big[\sum_{s=t}^T D_s|\mathcal{F}_t\big]\,, \ D\in\cD, \ A\in\cF_t.
\end{equation}
Combining \eqref{proof:RM-Representation-12-2} and \eqref{proof:RM-Representation-12-3}
dynamic consistency of $\{\cQ_t\}_{t=0}^T$ follows.

This completes the proof.
\endproof

\begin{remark}\label{rem:afterThDRM}
An interesting question is whether the sequence $\{\cQ_s\}_{s=0}^T$ appearing in the representation \eqref{eq:RM-Representation} can be replaced with a constant sequence of sets of probability measures. The question is motivated by the following observation: \\
First note that
for any set of probability measures $\mathcal{Q}\subseteq\mathcal{P}$, the following inequality holds true
\begin{align}
1_A\min_{\omega\in A}\bigg\{\inf_{\mathbb{Q}\in\mathcal{Q}}\mathbb{E}_{\mathbb{Q}}[X|\mathcal{F}_{t+1}](\omega)\bigg\}
& \leq 1_A\inf_{\mathbb{Q}\in\mathcal{Q}}\mathbb{E}_{\mathbb{Q}}[X|\mathcal{F}_{t}]\, , \label{eq:SetofProbs1}
\end{align}
for every $t\in\{0,\dots,T-1\}$, $A\in\mathcal{F}_t$,  and every random variable $X$.
Thus, if the set $\mathcal{Q}$ additionally satisfies the following {\em weak consistency} condition
\begin{equation}\label{eq:set-weak-consistency}
1_A\max_{\omega\in A}\bigg\{\inf\limits_{Q\in\mathcal{Q}} \mathbb{E}_{Q}\big[ X|\mathcal{F}_{t+1} \big](\omega)\bigg\} \geq
1_A\inf\limits_{Q\in\mathcal{Q}} \mathbb{E}_{Q}\big[ X |\mathcal{F}_{t} \big]\,,
\end{equation}
then the constant sequence $\mathcal{Q}_t= \mathcal{Q}, \ t\in\mathcal{T}$, is dynamically consistent. Observe that in Example \ref{ex:RM4} we indeed have that
$$
\rho_t(D)=-\inf_{\mathbb{Q}\in\mathcal{Q}}\mathbb{E}_{\mathbb{Q}}\big[\sum_{s=t}^T D_s|\mathcal{F}_t\big], \quad
 t\in\cT, \ D\in\mathcal{D},
$$
where $\mathcal{Q}=\mathcal{P}^e$. Note however that $\mathcal{P}^e$ satisfies consistency condition \eqref{eq:set-consistent} which is stronger than \eqref{eq:set-weak-consistency}.

\end{remark}

\subsection{Representation of DCAIs}
Having derived a representation theorem for dynamic coherent risk measures in terms of sets of probability measures, and having derived the duality between DCRM and DCAI
we can present the final results of this paper: duality between DCAI and sets of probability measures.

\begin{definition}
A family of sequences of sets of probability measures \\
$(\mathcal{U}^x:=(\mathcal{Q}^x_t)_{t=0}^T)_{x\in(0,+\infty)}$ is called increasing
if $\mathcal{Q}_t^x\supseteq \mathcal{Q}_t^y$, for all $x\geq y>0$ and $t\in\cT$.
\end{definition}
As direct consequence of Theorem \ref{th:RM-Representation} and Theorem  \ref{th:AI-Representedby-RM} we have the following results:

\begin{theorem}\label{th:Repr-DAI-1}
Assume that $(\mathcal{U}^x:=(\mathcal{Q}^x_t)_{t=0}^T)_{x\in(0,+\infty)}$ is
an increasing family of dynamically consistent sequences of sets of probability measures.
Then,  the function $\alpha: \{0,1,\ldots,T\}\times\mathcal{D}\times\Omega\to[0,+\infty]$ defined as follows,
\begin{equation}\label{eq:ReprDCAIbySetsOfProbs}
\alpha_t(D)=\sup\{x\in(0,+\infty): \inf_{\mathbb{Q}\in\mathcal{Q}^x_t}\mathbb{E}_{\mathbb{Q}}\big[\sum_{s=t}^T D_s|\mathcal{F}_t\big]\geq 0\} \,,
\quad t\in\cT, \ D\in\mathcal{D},
\end{equation}
is a normalized and right-continuous dynamic coherent acceptability index.
\end{theorem}

\begin{theorem}
If $\alpha$ is a normalized and right-continuous dynamic coherent acceptability index,
then there exists a family of dynamically consistent sequences of sets of probability measures
$(\mathcal{U}^x:=(\mathcal{Q}^x_t)_{t=0}^T)_{x\in(0,+\infty)}$ such that
\begin{equation*}
\alpha_t(D)=\sup\{x\in(0,+\infty): \inf_{\mathbb{Q}\in\mathcal{Q}^x_t}\mathbb{E}_{\mathbb{Q}}\big[\sum_{s=t}^T D_s|\mathcal{F}_t\big]\geq 0\} \,,
\quad t\in\cT, \ D\in\cD.
\end{equation*}
Here we adopt the usual convention that  $\inf\emptyset=\infty$ and $\sup\emptyset=0$.
\end{theorem}

\begin{remark}
We want to mention that the static AI is a particular case of the DCAI  developed in this paper and corresponds to $T=1$.
Same is true for the representation theorem for static AI in terms of family of sets of probability measures.
\end{remark}

\section{Examples}\label{sec:EX}
Theorem \ref{th:Repr-DAI-1}, besides being a fundamental theoretical result, can serve as basis for construction of DCAIs by means of constructing increasing sequences of dynamic sets of probability measures. Using this idea, we present here some abstract, non-trivial, examples of DCAIs.

\begin{example} {\bf Dynamic upper-limit ratio}. \\
Assume that $h:(0,+\infty)\to[0,+\infty)$ is an increasing function.
Define $\acute{\mathcal{Q}}^x$ as follows,
$$
\acute{\mathcal{Q}}^x:=\{\mathbb{Q}\in\mathcal{P}|\mathbb{E}_{\mathbb{P}}[{d\mathbb{Q}\over d\mathbb{P}}|\mathcal{F}_j]\leq (1+h(x))\mathbb{E}_{\mathbb{P}}[{d\mathbb{Q}\over d\mathbb{P}}|\mathcal{F}_{j-1}] \textrm{ for all } j=1,\dots,T, \}\,,
$$
and let  $\mathcal{U}^x:=\{\acute{\mathcal{Q}}^x\}_{t=0}^T$.
Note that $\acute{\mathcal{Q}}^x = \cQ^{1+h(x),u}, \ x\geq0$, where $\cQ^{a,u}, a\geq1$,
 is defined in Example \ref{ex:Set-Prob-3}, and
thus $\acute{\mathcal{Q}}^x$ is dynamically consistent for any $x>0$. Also observe that monotonicity of $h$ implies monotonicity of $\acute{\mathcal{Q}}^x$
with respect to $x$. Hence, by Theorem \ref{th:Repr-DAI-1},
\begin{equation*}
\alpha_t(D)=\sup\{x\in(0,+\infty): \inf_{\mathbb{Q}\in\acute{\mathcal{Q}}^x}\mathbb{E}_{\mathbb{Q}}\big[\sum_{s=t}^T D_s|\mathcal{F}_t\big]\geq 0\} \,.
\end{equation*}
is a normalized and right-continuous dynamic coherent acceptability index.
We call it {\it dynamic upper-limit ratio}.
\end{example}

\begin{example}\label{ex:RM4} {\bf Dynamic lower-limit ratio}. \\
Similarly, using Example \ref{ex:Set-Prob-4}, we consider
$\grave{\mathcal{Q}}^x:= \cQ^{1+h(x),l}$, for some increasing, non-negative function $h$.
Then, $\mathcal{U}^x:=\{\grave{\mathcal{Q}}^x\}_{t=0}^T$ is dynamically consistent,
and by Theorem \ref{th:Repr-DAI-1}, the function $\alpha$ defined by \eqref{eq:ReprDCAIbySetsOfProbs}
with $\cQ_t^x= \grave{\mathcal{Q}}^x, \ x>0$, is a normalized and right-continuous dynamic coherent acceptability index.
We call it {\it dynamic lower-limit ratio}.
\end{example}

\begin{example} (Continuation of Example \ref{ex:Set-Prob-5})

In Example \ref{ex:Set-Prob-5} we constructed a non-constant dynamically consistent sequence of sets of probability measures. In view of \eqref{eq:Ex-Prob-Sets5} the corresponding family of risk measures satisfies
$$
\rho_t(D)=-\inf_{\mathbb{Q}\in\mathcal{Q}_t}\mathbb{E}_{\mathbb{Q}}\big[\sum_{s=t}^T D_s|\mathcal{F}_t\big]
=-\inf_{\mathbb{Q}\in\mathcal{P}^e}\mathbb{E}_{\mathbb{Q}}\big[\sum_{s=t}^T D_s|\mathcal{F}_t\big] \, , \quad
\textrm{for all} \  t\in\cT, \ D\in\mathcal{D}.
$$
The point that we are making here is that the infimum over a time dependent set $\mathcal{Q}_t$ can be replaced by the infimum over time independent set $\mathcal{P}^e$ (see also Remark \ref{rem:afterThDRM} in this regard).
\end{example}

\begin{example} {\bf Dynamic Gain Loss Ratio.} \\
Gain Loss Ratio (GLR) is a typical return-to-risk type of performance measure, very popular among practitioners. We recall that it is defined as the ratio of expectation of positive returns to expectation of negative returns:
$\mathrm{GLR}(X):=\mathbb{E}(X)/\mathbb{E}(\max\{-X,0\})$, if $\mathbb{E}[X]>0$, and zero otherwise.
As shown in \cite{ChernyMadan2009}, GLR is a (static) coherent acceptability  measure.

 Here we present a dynamic version of GLR, denoted by dGLR, and defined as follows:
\begin{equation}\label{def:DLR}
\mathrm{dGLR}_t(D):=
\begin{cases}
\frac{\mathbb{E}[\sum_{s=t}^T D_s|\mathcal{F}_t]}
{\mathbb{E}[\left(\sum_{s=t}^T D_s \right)^{-}\,|\,\mathcal{F}_t] }, &
\quad \mathrm{if} \quad  \mathbb{E}[ \sum_{s=t}^T D_s \,|\,\mathcal{F}_t] >0\,,\\
0, & \quad \mathrm{otherwise}\,,
\end{cases}
\end{equation}
where $(\sum_{s=t}^T D_s)^{-}:=\max\{-\sum_{s=t}^T D_s,0\}$ and $t\in\cT, \ D\in\cD$.
Note that taking $T=1$, dGLR becomes the static GLR.
\end{example}

We argue that {\it dGLR is a normalized and right-continuous dynamic coherent acceptability index.}
Indeed, since $\textrm{dGLR}(1_{T})=+\infty$ and $\textrm{dGLR}(-1_{T})=0$, we have that dGLR is normalized.
Right-continuity follows from linearity of expectation and continuity of function $f(x)=x^{-}$.
Adaptiveness (D1), and independence of the past (D2) of dGLR follow directly from the definition.
Monotonicity (D3), scale invariance (D4), and quasi-concavity (D5) are verified as in static case with expectation replaced by
conditional expectation (for details see \cite{ChernyMadan2009}).

Since $\mathbb{E}(\sum_{l=t}^T (D+m1_{\{s\}})_l|\mathcal{F}_t) = \mathbb{E}(\sum_{l=t}^T (D+m1_{\{t\}})_l|\mathcal{F}_t)$,
and $\mathbb{E}( (\sum_{l=t}^T (D+m1_{\{s\}})_l)^- |\mathcal{F}_t) =
\mathbb{E}((\sum_{l=t}^T (D+m1_{\{t\}})_l)^-|\mathcal{F}_t)$, for all $t\in\cT, \ D\in\cD$,
(D6), translation invariance, follows.

Finally we will prove that dGLR satisfies (D7), dynamic consistency property.
Assume that $m$ is an $\mathcal{F}_t$-measurable random variable, and $D\in\mathcal{D}$ such that
$D_t\leq 0$ and $\textrm{dGLR}_{t+1}(D)\leq m$. Assume that $m\neq + \infty$, and $\mathbb{E}[\sum_{s=t+1}^T D_s|\mathcal{F}_{t+1}]>0$.
By definition of dGLR, we have,
$\mathbb{E}(\sum_{s=t+1}^TD_s|\mathcal{F}_{t+1})\leq m\cdot \mathbb{E}(\{\sum_{s=t+1}^T D_s\}^-|\mathcal{F}_{t+1})$, and since
$D_t\leq 0$, we have
\begin{align*}
\mathbb{E}(\sum_{s=t}^TD_s|\mathcal{F}_{t})
\leq\mathbb{E}(\mathbb{E}(\sum_{s=t+1}^TD_s|\mathcal{F}_{t+1})|\mathcal{F}_{t})
\leq m\mathbb{E}(\{\sum_{s=t+1}^T D_s\}^-|\mathcal{F}_t) \leq m \mathbb{E}(\{\sum_{s=t}^T D_s\}^-|\mathcal{F}_t) \, .
\end{align*}
which implies that $\textrm{dGLR}_t(D)\leq x$. If $m=+\infty$ or $\mathbb{E}[\sum_{s=t+1}^T D_s|\mathcal{F}_{t+1}]\leq 0$, then clearly $\textrm{dGLR}_t(D)\leq m$.

Similarly, one can show that if $D_t\geq 0$, and $\textrm{dGLR}_{t+1}(D)\geq m$, then $\textrm{dGLR}_t(D)\geq m$.

\noindent  Thus, we conclude that dGLR is a DCAI.

\bigskip

\begin{figure}[!htb]\label{fig:draroc}
\begin{center}
  \includegraphics[width=.6\textwidth]{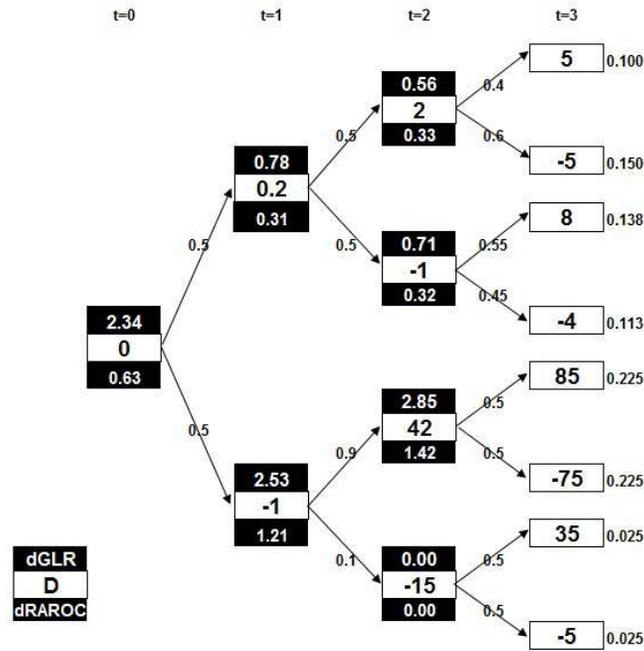}
  \caption{dRAROC vs dGLR}
\end{center}
\end{figure}

\begin{example}[Counterexample]

Taking into account the general  form of a dynamic acceptability index (cf. \eqref{eq:ReprDCAIbySetsOfProbs}),
and the general form of a static one (cf. \eqref{eq:repStaticCAI}), the natural question arises:
is it possible to dynamize a static coherent acceptability index by taking the appropriate `conditional quantity'
of the cumulative future cash-flow? For example, to dynamize GLR, we consider the static GLR, and
replaced in it the expectation with conditional expectation, and the terminal value with future cumulative cash-flow.
However, this procedure is not valid in general.
The natural extension of static Risk Adjusted Return on Capital (RAROC) to a dynamic setup has the following form:
\begin{equation*}
\textrm{dRAROC}_t(D)=
\begin{cases}
         {\mathbb{E}(\sum_{s=t}^T D_s|\mathcal{F}_t)\over
-\mathop{\inf}\limits_{\mathbb{Q}\in\mathcal{Q}}\mathbb{E}_{\mathbb{Q}}[\sum_{s=t}^T D_s|\mathcal{F}_t]}, & \text{when $\mathbb{E}(\sum_{s=t}^T D_s|\mathcal{F}_t) > 0$} \\
                  0, & \textrm{ otherwise}
\end{cases}
\end{equation*}
with convention $\textrm{dRAROC}_t(D)=+\infty$ if $\mathop{\inf}\limits_{\mathbb{Q}\in\mathcal{Q}}\mathbb{E}_{\mathbb{Q}}[\sum_{s=t}^T D_s|\mathcal{F}_t]\geq 0$.
\end{example}
As it is seen from Figure \ref{fig:draroc}, which represents  a numerical example, dRAROC does not satisfy property (D7), dynamic consistency.
In this example, we consider $\mathcal{Q}=\cP^e$. Assume that the states are labeled from top to bottom $\omega_1,\omega_2,  \ldots, \omega_8$.
Note that, $D_1(\omega_1)=0.2 >0$, i.e. positive cashflow at time $t=1$ and state $\omega_1$, but
$\textrm{dRAROC}_1(\omega_1) = 0.31 < 0.33 = \textrm{dRAROC}_2(\omega_1)$, as well as
$\textrm{dRAROC}_1(\omega_1) = 0.31 < 0.32 = \textrm{dRAROC}_2(\omega_2)$. Thus dRAROC does not satisfy (D7) and hence it is not a DCAI.

For comparison reasons, we also present in Figure \ref{fig:draroc} the values of dGLR, which is a DCAI.

\bibliographystyle{amsplain}

\begin{thebibliography}{10}

\bibitem{AcciaioFollmerPenner2010}
B.~Acciaio, H.~F\"{o}llmer, and I.~Penner, \emph{Risk assessment for uncertain
  cash flows: Model ambiguity, discounting ambiguity, and the role of bubbles},
  preprint (2010).

\bibitem{AcciaioPenner2010}
B.~Acciaio and I.~Penner, \emph{Dynamic risk measures}, preprint (2010).

\bibitem{ADEH1997}
P.~Artzner, F.~Delbaen, J.-M. Eber, and D.~Heath, \emph{Thinking coherently},
  Risk \textbf{10} (1997), 68--71.

\bibitem{ADEH1999}
\bysame, \emph{Coherent measures of risk}, Math. Finance \textbf{9} (1999),
  no.~3, 203--228.

\bibitem{ArtzerDelbaenEberHeathKu2007}
P.~Artzner, F.~Delbaen, J.-M. Eber, D.~Heath, and H.~Ku, \emph{Coherent
  multiperiod risk adjusted values and bellman's principle}, Ann. Oper. Res.
  \textbf{152} (2007), 5--22.

\bibitem{GLR2000}
A.~Bernardo and O.~Ledoit, \emph{Gain, loss, and asset pricing}, Journal of
  Political Economy \textbf{108} (2000), 144--172.

\bibitem{BionNadal2009}
J.~Bion-Nadal, \emph{Time consistent dynamic risk processes}, Stochastic
  Process. Appl. \textbf{119} (2009), no.~2, 633--654.

\bibitem{CarGemanMadan2001}
P.~Carr, H.~Geman, and D.B. Madan, \emph{Pricing and hedging in incomplete
  markets}, J. Fin. Econ. \textbf{62} (2001), no.~1, 131--167.

\bibitem{CheriditoDelbaenKupper2004}
P.~Cheridito, F.~Delbaen, and M.~Kupper, \emph{Coherent and convex monetary
  risk measures for bounded c\`adl\`ag processes}, Stochastic Process. Appl.
  \textbf{112} (2004), no.~1, 1--22.

\bibitem{CheriditoDelbaenKupper2005}
\bysame, \emph{Coherent and convex monetary risk measures for unbounded
  c\`adl\`ag processes}, Finance Stoch. \textbf{9} (2005), no.~3, 369--387.

\bibitem{CheriditoDelbaenKupper2006}
\bysame, \emph{Dynamic monetary risk measures for bounded discrete-time
  processes}, Electron. J. Probab. \textbf{11} (2006), no. 3, 57--106.

\bibitem{CheriditoLi2009}
P.~Cheridito and T.~Li, \emph{Risk measures on {O}rlicz hearts}, Math. Finance
  \textbf{19} (2009), no.~2, 189--214.

\bibitem{ChernyMadan2009}
A.S. Cherny and D.B. Madan, \emph{New measures for performance evaluation}, The
  Review of Financial Studies \textbf{22} (2009), no.~7, 2571--2606.

\bibitem{CvitanicKaratzas1999}
J.~Cvitanic and I.~Karatzas, \emph{On dynamic meaures of risk}, Finance Stoch.
  \textbf{3} (1999), no.~4, 451--482.

\bibitem{Delbaen2000}
F.~Delbaen, \emph{Coherent risk measures}, Scuola Normale Superiore, 2000.

\bibitem{Delbaen2002}
\bysame, \emph{Coherent risk measures on general probability spaces}, Advances
  in finance and stochastics, Springer, 2002, pp.~1--37.

\bibitem{DelbaenPengGianin2010}
F.~Delbaen, S.~Peng, and E.~Rosazza Gianin, \emph{Representation of the penalty
  term of dynamic concave utilities}, Finance and Stochastics \textbf{14}
  (2010), 449--472, 10.1007/s00780-009-0119-7.

\bibitem{Detlefsen2005}
K.~Detlefsen and G.~Scandolo, \emph{Conditional and dynamic convex risk
  measures}, Finance Stoch. \textbf{9} (2005), no.~4, 539--561.

\bibitem{FollmerSchied2002a}
H.~F{\"o}llmer and A.~Schied, \emph{Convex measures of risk and trading
  constraints}, Finance Stoch. \textbf{6} (2002), no.~4, 429--447.

\bibitem{FollmerSchied2002b}
\bysame, \emph{Robust preferences and convex measures of risk}, Advances in
  finance and stochastics, Springer, 2002, pp.~39--56.

\bibitem{FollmerSchiedBook2004}
H.~F{\"o}llmer and A.~Schied, \emph{Stochastic finance}, extended ed., de
  Gruyter Studies in Mathematics, vol.~27, Walter de Gruyter \& Co., Berlin,
  2004, An introduction in discrete time.

\bibitem{FrittelliGianin2002}
M.~Frittelli and E.~Rosazza~Gianin, \emph{Putting order in risk measures},
  Journal of Banking and Finance \textbf{26} (2002), 1473--1486.

\bibitem{FrittelliGianin2004}
\bysame, \emph{Dynamic convex measures}, 2004, pp.~227--248.

\bibitem{FrittelliGianin2005}
\bysame, \emph{Law invariant dynamic convex measures}, 2005, pp.~227--248.

\bibitem{FrittelliScandolo2006}
M.~Frittelli and G.~Scandolo, \emph{Risk measures and capital requirements for
  processes}, Math. Finance \textbf{16} (2006), no.~4, 589--612.

\bibitem{JobertRogers2006}
A.~Jobert and L.C.G. Rogers, \emph{Valuations and dynamic convex risk
  measures}, Math. Finance \textbf{18} (2008), no.~1, 1--22.

\bibitem{Kusuoka2001}
S.~Kusuoka, \emph{On law invariant coherent risk measures}, Advances in
  mathematical economics, {V}ol.\ 3, Adv. Math. Econ., vol.~3, Springer, 2001,
  pp.~83--95.

\bibitem{Riedel2004}
F.~Riedel, \emph{Dynamic coherent risk measures}, Stochastic Process. Appl.
  \textbf{112} (2004), no.~2, 185--200.

\bibitem{Roorda2005}
B.~Roorda, J.M. Schumacher, and J.~Engwerda, \emph{Coherent acceptability
  measures in multiperiod models}, Math. Finance \textbf{15} (2005), no.~4,
  589--612.

\bibitem{Sharpe1966}
W.F. Sharpe, \emph{Mutual fund performance}, Journal of Business \textbf{39}
  (1966), 119--139.

\bibitem{SortinoPrice1994}
F.A. Sortino and L.N. Price, \emph{Performance measurement in a downside risk
  framework}, The Journal of Investing \textbf{3} (1994), no.~3, 59--64.

\bibitem{Tutsch2008}
S.~Tutsch, \emph{Update rules for convex risk measures}, Quant. Finance
  \textbf{8} (2008), no.~8, 833--843.

\bibitem{Weber2006}
S.~Weber, \emph{Distribution-invariant risk measures, information, and dynamic
  consistency}, Math. Finance \textbf{16} (2006), no.~2, 419--441.

\end{thebibliography}

\providecommand{\bysame}{\leavevmode\hbox to3em{\hrulefill}\thinspace}
\providecommand{\href}[2]{#2}

\end{document}